\documentclass[traditabstract]{aa}

\usepackage{natbib}
\bibpunct{(}{)}{;}{a}{}{,} 
\usepackage{graphicx}

\newcommand{\cBE}{sph,cr}
\newcommand{\nrho}{\bar \mu m_{\rm H}}
\newcommand{\FWHM}{{\rm{fwhm}}}
\newcommand{\FWHMEQ}{{\it{fwhm}}}
\newcommand{\pindex}{\alpha}

\begin{document}

\title{Physical properties of interstellar filaments }

\author{J\"org Fischera\inst{\ref{inst1},}\inst{\ref{inst2}} and Peter G. Martin\inst{\ref{inst1}}}
\institute{Canadian Institute for Theoretical Astrophysics, University of Toronto, 60 St. George Street, ON M5S3H8, 
	Canada\label{inst1}
	\and
	Research School of Astronomy \& Astrophysics, 
	Institute of Advanced Studies, The Australian National University,
	Cotter Road, Weston Creek, ACT 2611 Australia \label{inst2}
	}


        \abstract{We analyze the physical parameters of interstellar
          filaments that we describe by an idealized model of
          isothermal self-gravitating infinite cylinder in pressure
          equilibrium with the ambient medium. Their gravitational
          state is characterized by the ratio $f_{\rm cyl}$ of their
          mass line density to the maximum possible value for a
          cylinder in a vacuum.  Equilibrium solutions exist only for
          $f_{\rm cyl} < 1$.  This ratio is used in providing
          analytical expressions for the central density, the radius,
          the profile of the column density, the column density
          through the cloud centre, and the \FWHM.  The dependence of
          the physical properties on external pressure and temperature
          is discussed and directly compared to the case of
          pressure-confined isothermal self-gravitating spheres.
          Comparison with recent observations of the \FWHM{} and the
          central column density $N_{\rm H}(0)$ show good agreement
          and suggest a filament temperature of $\sim 10~{\rm K}$ and
          an external pressure $p_{\rm ext}/k$ in the range $1.5\times 10^{4}~{\rm
            K~cm}^{-3}$ to $5\times 10^4~{\rm K~cm^{-3}}$.
          Stability considerations indicate that interstellar
          filaments become increasingly gravitationally unstable with
          mass line ratio $f_{\rm cyl}$ approaching unity.  For
          intermediate $f_{\rm cyl}>0.5$ the instabilities should
          promote core formation through compression, with a separation
          of about five times the \FWHM.
          We discuss the nature of filaments with high mass line
          densities and their relevance to gravitational fragmentation
          and star formation.}

\maketitle

\section{Introduction}

Filamentary structures are an ubiquitous phenomenon in the
interstellar medium.
Thanks to the high angular resolution and signal to noise of dust
imaging with the \emph{Herschel Space Observatory}, it is now possible
to quantify the basic empirical properties of filaments.
Filaments have been imaged in the submillimetre in exquisite detail in
non star-forming such as \object{Polaris} \citep{Mensch2010,mamd2010},
in molecular regions with low-mass star formation such as
\object{Aquila} and \object{IC 5146}
\citep{Andre2010,Arzoumanian2011,Mensch2010}, and in higher-mass
regions including \object{Vela C} and the \object{Rosette molecular cloud}
\citep{Hill2011,Schneider2012}.  
At least in the low-mass star forming regions and in \object{Polaris} ,
the characteristic filament sizes are about 0.1~pc
\citep{Arzoumanian2011}.\footnote{Appendix~\ref{App_uncertainty}
  discusses uncertainties in this and other important physical
  parameters like column density.}

Clear evidence for star formation seems to be related to cloud
structures with a typical extinction larger then $A_V\approx 7~{\rm
  mag}$ \citep{Onishi1998,Johnstone2004,Andre2010}, which corresponds
to a column density $N_{\rm H}$ about $0.6 \times 10^{22}$~cm$^{-2}$
(see comments in Appendix~\ref{App_uncertainty}).\footnote{According
  to \citet{Enoch2007}, however, such a threshold might vary from
  region to region.}  Furthermore, the \emph{Herschel} studies show
that high column densities are typically associated with filaments
and, finally, that cold dense clouds called ``prestellar cores''
possibly related to the very early stages of the star formation
\citep{Konyves2010}, are observed mainly along filaments
\citep{Andre2010,Mensch2010,Arzoumanian2011}.
Empirically, it is found that filaments that are star forming are
characterized by not only a higher central column density but also a
higher mass per unit length \citep{Andre2010,Arzoumanian2011}.  It is
argued that the cores are possibly created out of the elongated
structures through gravitational instabilities
\citep{Schneider1979,Gaida1984,Hanawa1993,Curry2000a,Andre2010,Mensch2010}.

It is important to develop a physical model for the filaments so that
the new measurements can be exploited.  That is the goal of this
paper.  
The basic model that we explore is an isothermal cylinder confined at
a finite boundary $r_{\rm cyl}$ by an external pressure $p_{\rm ext}$
provided by an ambient or embedding medium.  This is certainly more
realistic than an isothermal cylinder in vacuum, whose properties
(such as the radial profile) are not surprisingly in disagreement with
observations (e.g., \citealp{Mensch2010,Arzoumanian2011}).
Even these pressure-confined models are unlikely to describe what is a
very complex interstellar medium where the origin of the filamentary
structures is still a mystery, but the models can serve the same role
in developing our understanding that has been played by the
spherically-symmetric analog, the Bonnor-Ebert sphere
\citep{Ebert1955,Bonnor1956,Nagasawa1987,Inutsuka1992,Inutsuka1997,Curry2000,FiegePudritz2000a,Kandori2005,Fischera2008}.
Furthermore, we find that the models do actually provide a consistent
description of the sizes and column densities recently reported for
filaments and some insight into the instabilities relating to
filaments.

We analyze systematically the physical properties of pressure-confined
filaments, comparing and contrasting these results with the
corresponding properties for spheres.  We develop analytical solutions
for several observables.  Parts of the results, on which we build, can
be found in earlier work on the spectral energy distribution of
interstellar clouds (\citealp{Fischera2008} -- paper~I)  and on the
spectral energy distribution of condensed cores embedded in
pressurized elongated or spherical clouds (\citealp{Fischera2011} --
paper~II).

Our paper is organized as follows.
The physical model is presented in Section~\ref{Sect_model}, from which
properties are derived in Section~\ref{Sect_derived}.
Observed properties corresponding to size and column density are
compared with the model in Section~\ref{Sect_observations} with
remarkable agreement.
Structure formation due to instability of the filaments is discussed
in Section~\ref{Sect_structure}.  It seems plausible that dominant
protostellar cores arranged along filaments could arise from a
``compressive instability'' \citep{Nagasawa1987} in high column
density filaments.  Some other instabilities seem less relevant.
A summary and discussion is provided in Section~\ref{Sect_summary}.
Several appendices deal with the astronomical context and details of
some calculations.

\section{Physical model}\label{Sect_model}

We assume that the filaments can be described by a model of isothermal
self-gravitating infinitely long cylinders.  For the isothermal gas,
the relationship between gas pressure $p$ and gas density $\rho$ is
\begin{equation} 
\label{sound}
K=p /\rho = Tk/(\mu m_{\rm H}),
\end{equation}
where $k$ is the Boltzmann constant, $\mu$ the mean molecular weight
(2.36 for a molecular cloud with cosmic abundances), and $m_{\rm H}$
the mass of the hydrogen atom; the isothermal temperature $T$ is
considered to be an effective value related to both the thermal and
the turbulent motion of the gas (see Appendix~\ref{scaletemperature}).
If the effective temperature is equal to the thermal or kinetic
temperature then $K=c_{\rm s}^2$, the square of the sound speed.
Furthermore, as discussed in the introduction, the clouds are assumed
to be in pressure equilibrium with the surrounding medium of pressure
$p_{\rm ext}$.  This does not imply that the external medium is
isothermal or at the same temperature; it is simply a boundary
condition. 

\subsection{Cloud profile}

Isothermal self-gravitating infinitely long cylinders have a
well-known density or pressure profile given by (see
e.g. \citet{Stodolkiewicz1963,Ostriker1964})
\begin{equation}
	\label{eq_densityprofile}
	\rho(r) = \frac{\rho_c}{(1+  (r/\sqrt{8}r_0)^2)  )^2},
\end{equation}
where 
\begin{equation}
	\label{r0def}
r_0^2 = K/(4\pi G \rho_c),
\end{equation} 
$\rho_c$ is the central density, and $G$ the gravitational constant.
For pressure-confined clouds the profile terminates where the cloud
pressure {is equal to the external pressure $p_{\rm ext}$}; this
defines the cylinder radius $r_{\rm cyl}$. As in papers I and II,
  in the following we refer to the ratio $p_{\rm c}/p_{\rm ext}$ as
  the ``overpressure.''

\subsection{Mass line density}\label{massline}

Analogous to the cloud mass in the case of spherical clouds we have in
the case of cylinders the mass per cloud length (mass line density),
given by
\begin{eqnarray}
	\label{eq_masslinedensity}
	\frac{M}{l}(r_{\rm cyl}) &=& \int_0^{r_{\rm cyl}}{\rm
          d}r\,2\pi r\,\rho(r) \nonumber \\
		&=&\frac{2K}{G}\left\{1-\frac{1}{1+{r_{\rm cyl}}^2/(8r_0^2)}\right\}.
\end{eqnarray}
In the limit $r_{\rm cyl}\gg \sqrt{8}r_0$ the mass per cloud
lengths approaches asymptotically a \emph{maximum} value given by
\begin{equation}
	\label{eq_maxmasslinedensity}
	\left(\frac{M}{l}\right)_{\rm max} = \frac{2K}{G} = 16.4\, \left(\frac{T}{10~{\rm K}}\right)\, {\rm M}_\odot~{\rm pc}^{-1}.
      \end{equation}
For pressurized clouds this gravitational state corresponds to an
infinite overpressure, but there is no dependence on $p_{\rm ext}$.

For comparison, in the case of an isothermal self-gravitating
  pressurized sphere, the maximum mass is identical to its
  \emph{critical} mass, given by (see Appendix~\ref{app1})
\begin{equation}
	M_{\rm \cBE} \approx 4.191 \frac{K^2}{\sqrt{4\pi G^3 p_{\rm
              ext}}}.
\label{mbe}
\end{equation}
This gravitational state is characterized by an overpressure $p_{\rm
  c}/p_{\rm ext}\approx 14.04$.  As shown in paper~II, a critical
stable state is related to the maximum possible pressure that a sphere
of given mass and $K$ (temperature) can produce at the cloud outskirts
under compression.  If the external pressure is increased slightly,
the resulting equilibrium configuration is unstable to gravitational
collapse, and so this is aptly called a ``critical stable sphere''.
The spherical clouds can be characterized as ``subcritical'' or
``supercritical'' depending on whether the overpressure is lower or
higher than the value for a critical stable sphere.  Note that for
overpressures higher than 14.04, the mass of an equilibrium
configuration is \emph{smaller} than the critical mass.  Thus we can
also say that for given $p_{\rm ext}$ and $K$ the critical mass is the
maximum mass for which an equilibrium solution exists.

Unlike $M_{\rm \cBE}$, the maximum mass line density depends only
linearly (compared to quadratically) on $K$ or cloud temperature and
does not depend at all on the size of the external pressure confining
the cloud.  We discuss the basic consequences in
Sect.~\ref{sect_pressure} and Sect.~\ref{sect_temperature}.

There are no equilibrium solutions for cylinders or spheres with
  masses above the corresponding maximum values \citep{McCrea1957}.
  They would collapse radially towards a line (spindle) or to a
  singular point, respectively.  This has led to the terminology
  ``supercritical mass'' when $M > M_{\rm \cBE}$, or ``supercritical
  filament'' when $M/l > (M/l)_{\rm max}$.  The gravitational states
  considered in this paper are all equilibrium solutions.  We prefer
  to reserve the term ``supercritical'' to describe the equilibrium
  states for spheres, as above, and note the fundamental difference
  that there are no such supercritical equilibrium states for
  cylinders (Sect.~\ref{sect_pressure} ). 


\subsubsection{Mass fraction $f_{\rm cyl}$}

A very useful quantity is the mass fraction
\begin{equation}
f_{\rm cyl}=(M/l)/(M/l)_{\rm max}\le 1.
\label{ratiof}
\end{equation}
For cylinders we find directly from Eqs.~\ref{sound} to \ref{ratiof}
that the overpressure
\begin{equation}
	\label{eq_centralpressure}
        p_{\rm c}/p_{\rm ext} = 1/(1 - f_{\rm cyl})^{2}.
\end{equation}
We can therefore use $f_{\rm cyl}$ to characterize the gravitational
state of the cylinders.  This mass ratio therefore appears in the
formulae below for the derived properties of cylinders in different
gravitational states.  Two instructive limits to consider are the case
of high overpressure or $f_{\rm cyl} \rightarrow 1$, which we shall
see corresponds to a narrow high column density filament with strong
self-gravity, and the case of vanishing overpessure or $f_{\rm cyl}
\rightarrow 0$, corresponding to a filament much more difficult to
recognize or characterize.  There is, however, no distinction between
``pressure-confined'' and ``self-gravitating'' cylinders, since both
features are intrinsic to all equilibrium models.

This parallels the approach used in paper~I where the gravitational
state of pressurized spheres was described by the mass ratio $f_{\rm
  sph}=M/M_{\rm \cBE}$ of the cloud mass relative to the mass of a
critical stable sphere.  While in paper~I the physical parameters for
spheres were discussed only for the physically stable regime
with overpressures below the critical value 14.04, we consider here
the whole physical range.  This leads to the curious situation
  that for large overpressures above the critical value the
  gravitational state is described by $f_{\rm sph}<1$.

\section{Derived properties}\label{Sect_derived}

In the following we describe the physical properties of molecular
cylinders and compare spheres for a given cloud temperature and
external pressure.  We adopt illustrative reference values for
temperature and pressure.  See Appendix~\ref{App_uncertainty} for
further discussion.  For the cold molecular clouds we adopt $K$
corresponding to 10~K ($c_{\rm s }= 0.19$~km~s$^{-1}$).  For the
external pressure we adopt $p_{\rm ext}/k=2\times 10^4~{\rm
  K~cm}^{-3}$, consistent with paper~I and paper~II.

\subsection{Central density}

\begin{figure}[htb]
	\includegraphics[width=\hsize]{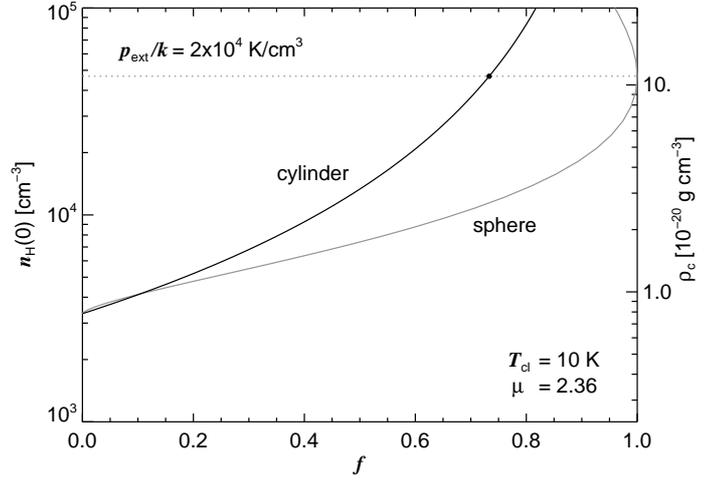}
	\caption{\label{fig_centraldensity} Central density of an
          isothermal self-gravitating molecular cylinder (black) and
          sphere (grey), for fixed values of the ambient pressure.
          $p_{\rm ext}/k=2\times 10^4~{\rm K~cm}^{-3}$ and cloud
          temperature $T = 10~{\rm K}$.  The dotted line is the value
          achieved by a critical stable sphere.  This same central
          density (or overpressure) is reached in a cylinder with
          $f_{\rm cyl}=0.733$ (filled black circle). }
\end{figure}
For the central density we have from
Eqs.~\ref{sound} and \ref{eq_centralpressure}
\begin{equation}
	\label{eq_centraldensity}
	\rho_{\rm c} = \frac{p_{\rm ext}}{K(1-f_{\rm cyl})^2}.
\end{equation}
The central number density $n_{\rm H} (r=0) \equiv n_{\rm H}(0)$ is
obtained by evaluating
\begin{equation}
	n_{\rm H} = \rho \left(\sum\frac{n_i}{n_{\rm H}}\mu_i m_{\rm
            H}\right)^{-1}  \equiv \rho \, (\nrho)^{-1},
\end{equation}
where $n_i/n_{\rm H}$ is the relative abundance of element $i$
compared to H, and $\mu_i$ is its relative mass units of $m_{\rm H}$;
on the right, the effective mass per H, $\bar \mu$, is 1.4 for cosmic
abundances. Figure~\ref{fig_centraldensity} shows these central values
as a function of the mass fraction.

For a given temperature and external pressure the central density
increases as $1/(1-f_{\rm cyl})^2$ and so in the limit $f_{\rm
  cyl}\rightarrow 1$ the density becomes infinite.

For a given mass fraction, the density in the cloud centre is
proportional to $p_{\rm ext}\mu/T$.  Cylinders have a higher
overpressure or overdensity in the cloud centre compared to spheres
(Fig.~\ref{fig_centraldensity}) for the same mass fraction, except in
the case of small mass fractions ($f<\approx 0.15$).  In cylinders the
overpressure of 14.04 corresponding to critical stable spheres is
achieved for a mass fraction of $f_{\rm cyl}\approx 0.733$.

\subsection{Cloud radius}

\begin{figure}[htbp]
	\includegraphics[width=\hsize]{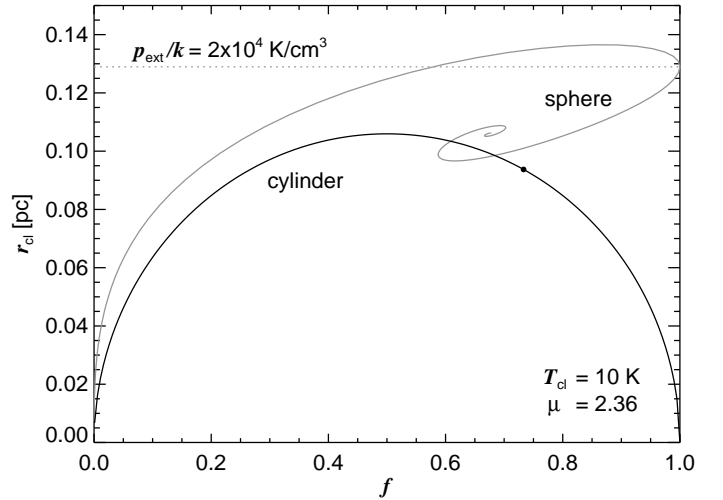}
	\caption{\label{fig_radius}Like Fig.~\ref{fig_centraldensity},
          but for the radius of an isothermal self-gravitating
          cylinder (black) and sphere (grey). }
\end{figure}

Solving Eq.~\ref{eq_masslinedensity} for $r_{\rm cl}$, using
Eq.~\ref{eq_centraldensity} and recasting Eq.~\ref{r0def}
as
\begin{equation}
\label{r0def1}
	r_0 = \frac{K}{\sqrt{4 \pi G p_{\rm ext}}} (1 - f_{\rm cyl}),
\end{equation}
provides
\begin{equation}
	\label{eq_cloudradius}
	r_{\rm cyl} = \frac{\sqrt{8}K}{\sqrt{4\pi G p_{\rm ext}}} \sqrt{\frac{1}{4}-\left(f_{\rm cyl}-\frac{1}{2}\right)^2}.
\end{equation}
This along with the boundary radius of the pressure-confined sphere is
plotted as $r_{\rm cl}$ in Figure~\ref{fig_radius} as a function of mass
fraction.

The size of the cloud for given mass fraction $f$ is proportional to
$(T/\mu)/\sqrt{p_{\rm ext}}$.  The radius for cylinders is symmetrical
around $f=0.5$ where the cloud shows its maximum size.  The filament
grows in size through accretion of new material as long as the mass
ratio $f_{\rm cyl}<0.5$.  Further accretion will cause a shrinking of
the filament.  In the limit $f_{\rm cyl}\rightarrow 1$ (the mass-line
density approaches the maximum value $2K/G$) the cylinder will become
infinitely thin.

Spheres by contrast have a finite size at large mass ratios. As
pointed out in paper~I the spherical cloud has a maximum size at
$f_{\rm sph}\approx 0.870$. Pushed above this value by accretion an
isothermal cloud size shrinks to the size of a critical stable
cloud. Supercritical stable spheres have smaller sizes still.

\subsection{Radial profile of the pressure or density}\label{radialdensity}

\begin{figure}[htbp]
	\includegraphics[width=\hsize]{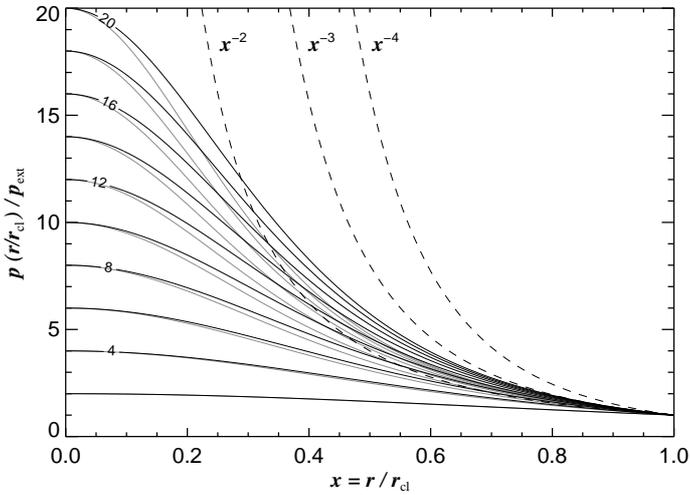}
	\caption{\label{figdensprofile} Pressure profiles of
          isothermal self-gravitating cylinders (black) and spheres
          (grey), for various overpressures $p_{\rm c}/p_{\rm ext}$
          from 2 to 20. The dashed curves show density profiles for
          simple power laws normalized at the cloud boundary.}
\end{figure}
 
The radial dependence of pressure given by
Eq.~\ref{eq_densityprofile} is shown in Fig.~\ref{figdensprofile}
for various overpressures, which according to
Eq.~\ref{eq_centraldensity} are equivalent to $1/(1-f_{\rm
  cyl})^2$.  For comparison, profiles are also shown for spheres for
the same set of overpressures; in the central region of the cloud
their density falls off more strongly towards larger radii.

In the limit of $r^2/(8r_0^2)\gg 2$ the clouds show a steep density
profile at the cloud edges with $\rho(r) \propto r^{-4}$. However,
this property requires clouds with very high overpressures (large
$f_{\rm cyl}$, close to the maximum possible mass-line density).  n
the outer half of the cylinders the density profiles for overpressures
less than 20 are shallower than a $r^{-4}$ profile. For overpressures
in the regime 6 to 12 the density profile at the outer half is more
consistent with a $r^{-2}$-profile. Compared to spheres,

A useful characteristic size for describing properties of the cylinder
is $r_{\rm cyl}$ for small $f_{\rm cyl}$ and $\sqrt{8} r_0$ for large
$f_{\rm cyl}$.

\subsection{Projection of the density profile}

\subsubsection{Column density profile}

The column density at an impact parameter $x$ (in units of cloud
radius $r_{\rm cl}$) is given by the integral
\begin{equation}
	N_{\rm H}(x) = 2\int\limits_0^{r_{\rm
            cyl}\sqrt{1-x^2}}\frac{n_{\rm H}(0)\,{\rm d} r}{(1+x^2 r_{\rm cyl}^2/8r_0^2+r^2/8 r_0^2)^2}.
\end{equation}
The integral evaluates to:
\begin{equation}
	N_{\rm H}(x) = n_{\rm H}(0)\sqrt{8}\frac{r_0}{c}\Big\{\frac{u}{u^2+c}+\frac{1}{\sqrt{c}}\arctan{\frac{u}{\sqrt{c}}}\Big\},
\end{equation}
with $c = 1+x^2 r_{\rm cyl}^2/8r_0^2$ and $u = \sqrt{1-x^2}\, r_{\rm cyl}/(\sqrt{8}r_0)$. 
Using the replacement $r_{\rm cyl}^2/8 r_0^2 = f_{\rm cyl}/(1-f_{\rm cyl})$ it is straightforward
to show that the column density profile for given $f_{\rm cyl}$ is
given by:
\begin{eqnarray}
	\label{eq_nhprofile}
	N_{\rm H}(x) &=& \sqrt{\frac{p_{\rm ext}}{4\pi G}}\frac{\sqrt{8}}{1-f_{\rm cyl}}
			(\nrho)^{-1}\nonumber\\
			 &&\times\frac{1-f_{\rm cyl}}{1-f_{\rm cyl}+x^2 f_{\rm cyl}}\Bigg\{
			 	\sqrt{f_{\rm cyl}(1-f_{\rm cyl})(1-x^2)}\nonumber\\
			&&+\sqrt{\frac{1-f_{\rm cyl}}{1-f_{\rm cyl}(1-x^2)}}\nonumber\\
			&&\times\arctan{\sqrt{\frac{f_{\rm cyl}(1-x^2)}{1-f_{\rm cyl}(1-x^2)}}}\Bigg\}.
\end{eqnarray}
The profile shape is solely determined by the mass fraction $f_{\rm
  cyl}$.  The amplitude of the profile with fixed $f_{\rm cyl}$ is
proportional to $\sqrt{p_{\rm ext}}$ and is independent of the cloud
temperature.

In the limit $f_{\rm cyl}\rightarrow 1$ the profile for $x\ll 1$ can
be approximated by:
\begin{eqnarray}
	\label{eq_nhprofapprox}
	N_{\rm H}(x ) &\approx& \sqrt{\frac{p_{\rm ext}}{4\pi G}}\frac{\sqrt{8}}{1-f_{\rm cyl}}
			 (\nrho)^{-1}\nonumber\\
			 &&\times\left(\frac{1-f_{\rm cyl}}{1-f_{\rm cyl}(1-x^2)}\right)^{3/2}\frac{\pi}{2}.
\end{eqnarray}

In the limit of low $f_{\rm cyl}$
\begin{equation}
	\label{eq_columnapprox0}
	N_{\rm H}(x)\approx (p_{\rm ext}/K) (\nrho)^{-1} 2 \sqrt{1-x^2}\, r_{\rm cyl},
\end{equation}
as for a uniform-density cylinder.

\begin{figure*}
	\includegraphics[width=0.499\hsize]{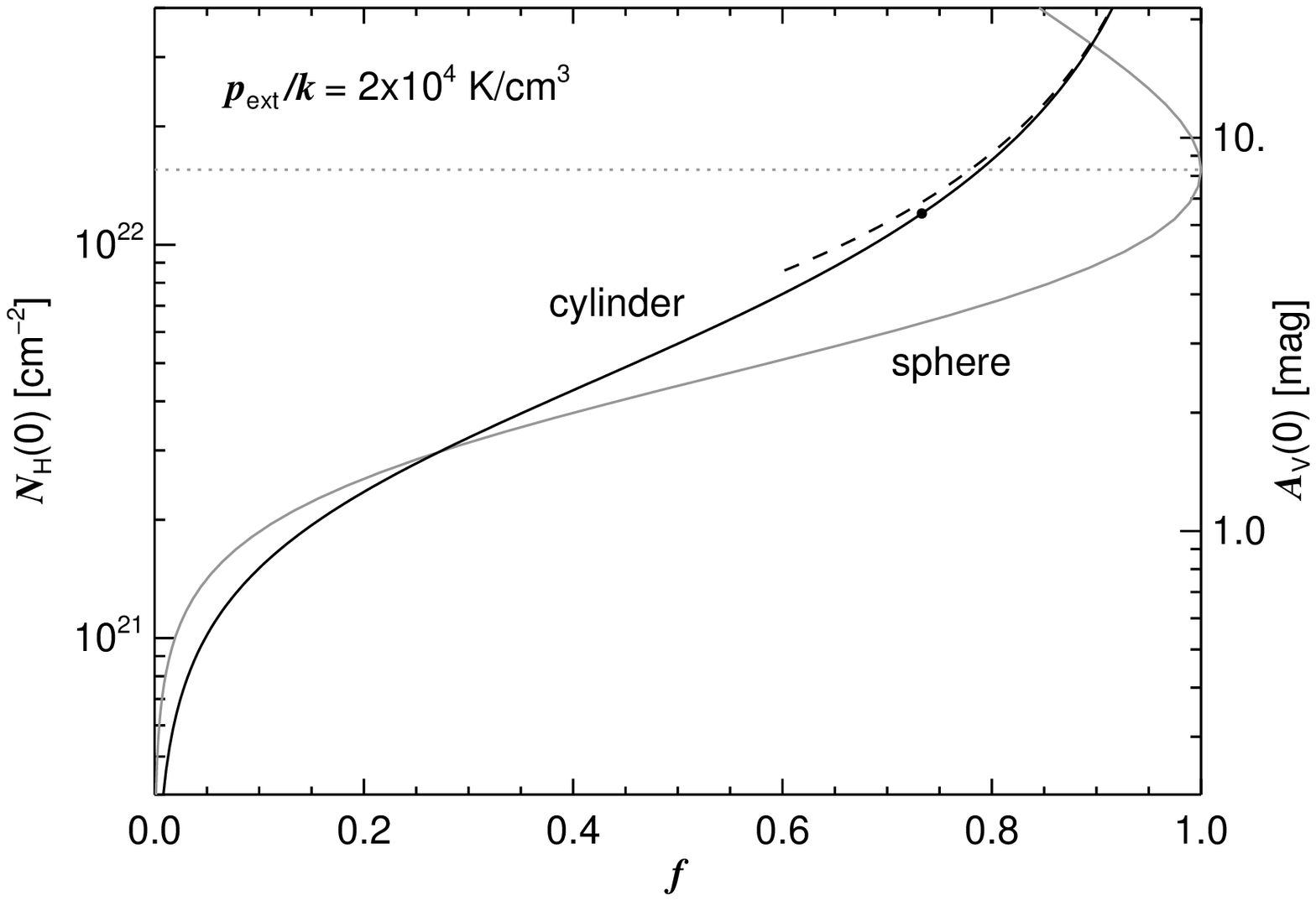}
	\hfill
	\includegraphics[width=0.499\hsize]{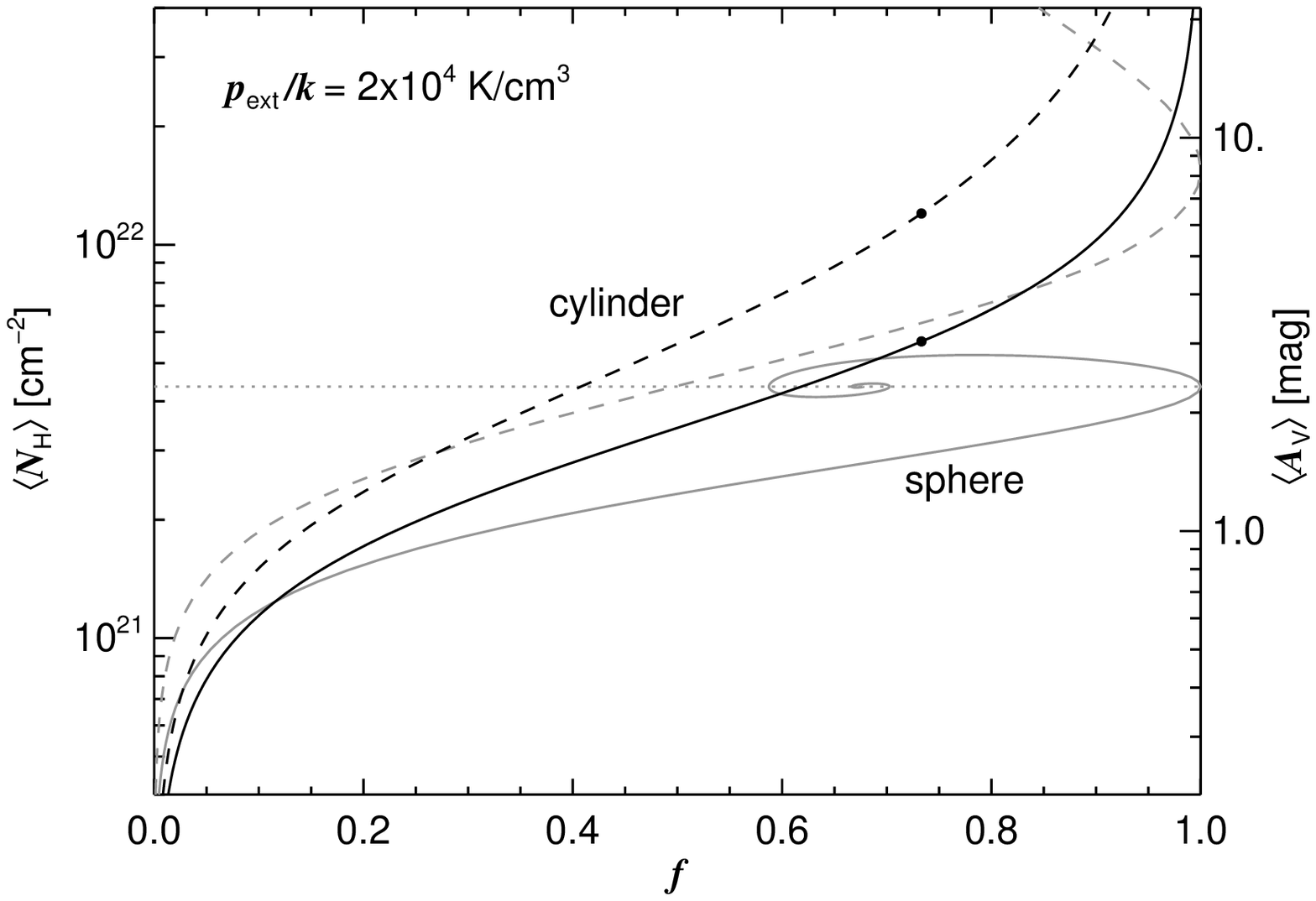}\\
	\caption{\label{fig_columndensity} Like
          Fig.~\ref{fig_centraldensity}, but for the central column
          density (left panel) and mean column density (right panel)
          through an isothermal self-gravitating cylinder (black) and
          sphere (grey), as a function of mass fraction $f$.  The
          dashed line in the left-hand plot is the approximation given
          in Eq.~\ref{eq_columnapprox}.  The dashed lines in the
          right-hand plot echo the central column densities from the
          left-hand plot for comparison.  The horizontal dotted lines
          give the values achieved by a critical stable sphere.  The
          values for a cylinder with that overpressure, mass fraction
          0.733, are shown as filled black circles.}
\end{figure*}

\subsubsection{\label{sect_columndensity}Central column density}

An important characteristic of self-gravitating clouds is the column
density through the cloud centre, $N_{\rm H}(0)$, which for cylinders
is a trivial solution of Eq.~\ref{eq_nhprofile} for $x=0$:
\begin{eqnarray}
	N_{\rm H}(0) &= & \sqrt{\frac{p_{\rm ext}}{4\pi G}}\frac{\sqrt{8}}{1-f_{\rm cyl}}(\nrho)^{-1}\nonumber\\
			&&\times\left\{\sqrt{f_{\rm cyl}(1-f_{\rm
                              cyl})}+\arctan\sqrt{\frac{f_{\rm
                                cyl}}{1-f_{\rm cyl}}}\right\}.
\label{eq:nh0}
\end{eqnarray}
Fig.~\ref{fig_columndensity} shows that the column density
increases more strongly with mass fraction in the case of cylinders.

In the limit of high mass fraction ($f_{\rm cyl}>\approx0.6$) the
column density can be approximated by
\begin{equation}
	\label{eq_columnapprox}
	N_{\rm H}(0)\approx
\sqrt{\frac{p_{\rm ext}}{4\pi G}}\frac{\sqrt{8}}{1-f_{\rm
    cyl}}(\nrho)^{-1}\frac{\pi}{2} = n_{\rm H}(0) \sqrt{8} r_0 \frac{\pi}{2}.
\end{equation}
For given external pressure the central column density increases as
$1/(1-f_{\rm cyl})$. In the limit of maximum mass fraction the column
density would be infinite. If we replace the mass fraction through the
overpressure we find that the column density for high overpressure
increases, as found in paper~II, proportional to $\sqrt{p_{\rm c}}$.
The limit for low $f_{\rm cyl}$ follows from Eq.~\ref{eq_columnapprox0}.

Column densities in the molecular environments in question are
measured via near-infrared colour excess or, in the case of the
filaments being discussed, the submillimetre optical depth
(Appendix~\ref{App_uncertainty}).  Nevertheless, $A_V$ is widely
used as ``shorthand'' for column density, which we do here as well,
even though it is not a direct observable in dense molecular clouds.
A column density of H nucleons of $10^{22}~{\rm cm^{-2}} $ is taken to
correspond to $A_V = 5.3$~mag
(Appendix~\ref{App_uncertainty}).\footnote{$N_{\rm H2}/(10^{22}~{\rm
    cm^{-2}} )$ is therefore approximately $A_V$.}
In other units, this is equivalent to $2.3 \times
10^{-2}$~gm~cm$^{-2}$ or $1.1 \times 10^{2} M_\odot$~pc$^{-2}$.  This
conversion relates the left and right vertical scales in
Fig.~\ref{fig_columndensity}.

For the adopted external pressure, the column density lies
between $2\times 10^{21}$ and $2\times 10^{22}~{\rm cm^{-2}}$ ($A_V$
in the range $1$ to $10$~mag) for mass fractions $f_{\rm cyl}$ in the
range roughly 0.2 to 0.8.

Cylinders with the same overpressure $p_{\rm c}/p_{\rm ext}$ as
spheres are characterized by a lower central extinction (paper~II).
For example, for the adopted external pressure, the central extinction
through a cylinder with the same overpressure as a critical stable
sphere is approximately $A_V\approx 6.4$~mag compared to 8~mag for
that sphere (Fig.~\ref{fig_columndensity}).

\subsubsection{\label{sect_avcolumndensity}Average column density}

For unresolved filaments we also give the mean column density (see
also paper~II). The mean value for cylinders is given by
\begin{eqnarray}
	\left<N_{\rm H}\right> &=& (\nrho)^{-1}\frac{M/l}{2r_{\rm
            cyl}} \nonumber \\
		&=& (\nrho)^{-1} \sqrt{\frac{4 \pi\, p_{\rm ext}}{8\, G}} \sqrt{\frac{f_{\rm cyl}}{1-f_{\rm cyl}}}.
\end{eqnarray}
We see that the mean column density of cylinders also approaches
infinity asymptotically in the limit of large mass ratio.

By contrast, the mean column density even for a supercritical sphere is
finite.  In the limit of large mass fraction the mean value is given
given by
\begin{eqnarray}
	\label{eq_meancolumn}
	\left<N_{\rm H}\right> &= & (\nrho)^{-1}\frac{M}{\pi r^2_{\rm
            sph}} \nonumber \\
		&\approx& (\nrho)^{-1} \sqrt{\frac{4 \pi\, p_{\rm ext}}{\pi^2\, G}} \sqrt{2}.
\end{eqnarray}
As shown in the App.~\ref{app1}, the value is exact where a cloud of
given mass produces a pressure maximum at the clouds outskirts.  The
value is for example exact for the special case of a critical stable
sphere or in the limit of infinite overpressure.  For $p_{\rm ext}/k=2\times 10^4~{\rm K~cm^{-3}}$
we have $\left<N_{\rm H}\right>\approx
4.36\times 10^{21}~{\rm cm^{-2}}$ ($A_V\approx 2.33~{\rm mag}$).  As
shown in Fig.~\ref{fig_columndensity} stable spherical clouds above
$f_{\rm sph}>0.1$ have mean column densities not smaller than
$10^{21}~{\rm cm^{-2}}$ and all spherical clouds, even if we include
the supercritical cases, mean column densities not larger than
$6\times 10^{21}~{\rm cm^{-2}}$.

\subsubsection{Column density profiles for different central $A_V$}

According to the previous sections, for a given external pressure the
column density or extinction in these interstellar clouds can be
specified by their overpressure or $f$, without explicit specification
of either the cloud temperature or their mass line density (filaments)
or mass (spheres).  For filaments the extinction is related
monotonically to the mass ratio $f_{\rm cyl}$.  The same applies to
spherical clouds up to overpressures $p_{\rm c}/p_{\rm ext}\approx
14.04$.  The extinction varies as $\sqrt{p_{\rm ext}}$.

\begin{figure}[htbp]
	\includegraphics[width=\hsize]{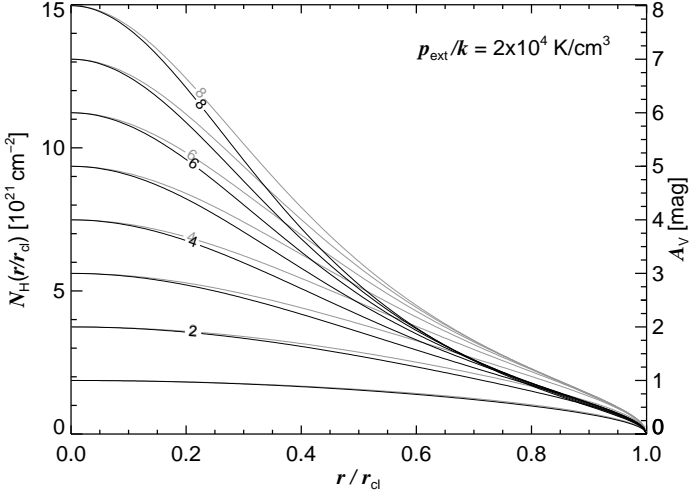}	
	\caption{\label{figcolumnprofile} Column density profiles of
          isothermal self-gravitating cylinders (black lines) and
          spheres (grey lines) for several central extinction values
          from $A_V=1~{\rm mag}$ to $A_V=8~{\rm mag}$.  The clouds are
          pressurized by $p_{\rm ext}/k=2\times
          10^4~{\rm K~cm}^{-3}$.  }
\end{figure}
\begin{figure}[htbp]
	\includegraphics[width=\hsize]{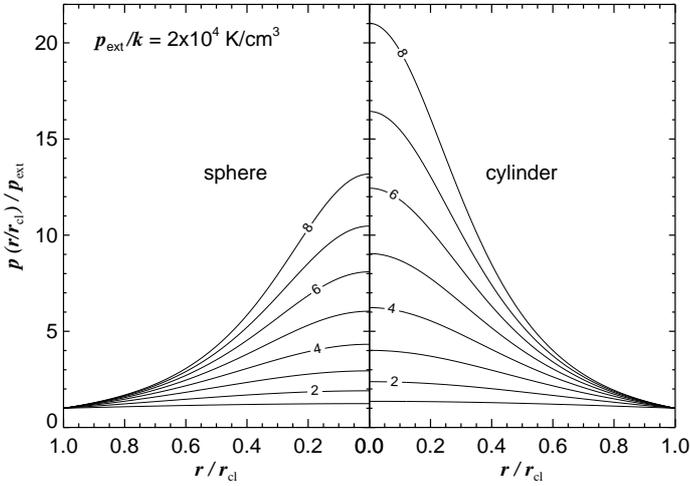}
	\caption{\label{figdensprofile2} Pressure profiles (density
          profiles) for the column density profiles shown in
          Fig.~\ref{figcolumnprofile}.  Recall that profiles for the
          same overpressure are compared in
          Fig.~\ref{figdensprofile}.}
\end{figure}

In Fig.~\ref{figcolumnprofile}, for the adopted $p_{\rm ext}/k=2\times
10^4~{\rm K~cm}^{-3}$, we give the profiles of the column densities of
cylinders and spheres for a number of different values of the central
extinction.  The column density profile steepens as a function of the
central extinction or $f$.  For the same central extinction, the
column density of a cylinder falls off more quickly with impact
parameter $x$ compared to a sphere.  This relative behaviour is
opposite that for the pressure profile (Fig.~\ref{figdensprofile}).

For the same central extinction, cylinders have a higher overpressure
$p_{\rm c}/p_{\rm ext}$ compared to spheres, as discussed in
Sect.~\ref{sect_columndensity} and shown in
Fig.~\ref{figdensprofile2}.

\subsection{The fwhm}
\label{sect_fwhm}

\begin{figure}[htbp]
	\includegraphics[width=\hsize]{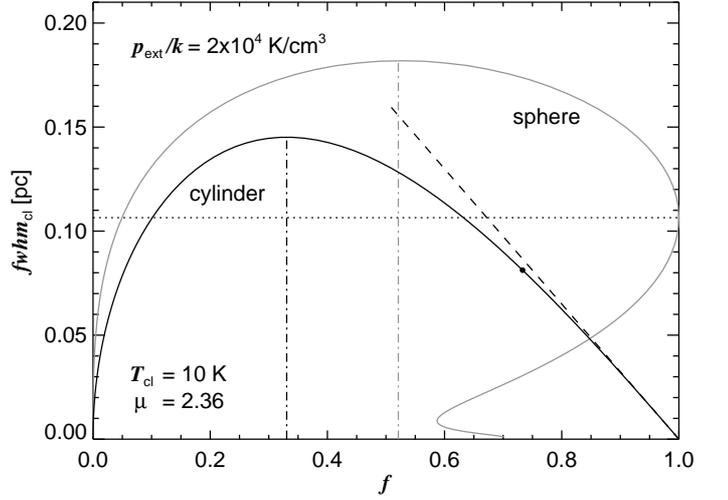}
	\caption{\label{fig_fwhm} Like Fig.~\ref{fig_radius}, but for
          the \FWHM{} of an isothermal self-gravitating cylinder (black
          line) and sphere (grey line).  The dashed line gives the
          asymptotic behaviour of the \FWHM{} of an
          isothermal self-gravitating cylinder for high $f$
          (Eq.~\ref{eq_fwhmapprox}).  The vertical dashed-dotted
          lines mark the mass fraction of the maxima of \FWHM{}
          (Eq.~\ref{eq_fwhmmax}).  The horizontal dotted line
          shows the \FWHM{} of a critical stable sphere.
          The \FWHM{} of a cylinder with this overpressure
          14.04 ($f_{\rm cyl}=0.733$) is shown as the filled black
          circle ($\FWHMEQ_{\rm cyl}$ = 0.0812~pc).  }
\end{figure}

Observationally the \FWHM{} has more relevance than the physical radius
which is either not directly observable or difficult to determine
because of the fluctuating background.  The impact parameter $x_{\FWHMEQ}$ 
of the \FWHM{} is given by
\begin{equation}
	N_{\rm H}(x_{\FWHMEQ}) = \frac{1}{2}N_{\rm H}(0).
\end{equation}
The \FWHM{} of cylinders and spheres is examined in
Appendix~\ref{sect_fwhmspheres}. For the adopted external pressure,
values are shown in Fig.~\ref{fig_fwhm}.

For the limit $f_{\rm cyl}\rightarrow 1$ an asymptotic behaviour of the
\FWHM{} of the cylinder can be found.  Using the approximation of the
column density profile (Eq.~\ref{eq_nhprofapprox}) leads to
\begin{equation}
	\label{eq_fwhmapprox}
	\FWHMEQ_{\rm cyl} 
	\approx 2 \sqrt{2^{2/3}-1} \sqrt{8} r_0.
\end{equation}
which decreases linearly as $(1-f_{\rm cyl})$.  As seen in the
Fig.~\ref{fig_fwhm} this behaviour is valid for $f_{\rm cyl}>\sim
0.7$. If we replace the mass ratio by Eq.~\ref{eq_centraldensity}
we see that the \FWHM{} decreases inversely as the square root of its
overpressure $p_{\rm c}/p_{\rm ext}$. The same behaviour is valid for
spheres (see App.~\ref{sect_fwhmspheres}).

For the limit $f_{\rm cyl}\rightarrow 0$ the asymptotic behaviour is
\begin{equation}
	\label{eq_fwhmapprox0}
	\FWHMEQ_{\rm cyl} = \sqrt{3}\, r_{\rm cyl}, 
\end{equation}
which varies as $\sqrt{f_{\rm cyl}}$.

In between these limits, the \FWHM{} shows a maximum size for both
spheres and cylinders, just as seen for $r_{\rm cl}$.  Because of the
steepening shape of the column density profile towards higher $f$ the
maxima appear at lower $f$ then that for the maximum $r_{\rm cl}$. The
maxima are given by
\begin{eqnarray}\label{eq_fwhmmax}
	\FWHMEQ_{\rm max} &=& \left(\frac{T}{10~{\rm K}}\right)\left(\frac{p_{\rm ext}/k}{2\times 10^4~{\rm K~cm}^{-3}}\right)^{-\frac{1}{2}}
		\nonumber\\
		&&\times  \left\{
\begin{array}{cc}
		0.145\,  {\rm pc}, & f_{\rm cyl} = 0.331 \\
		0.181\,  {\rm pc}, & f_{\rm sph} = 0.521
\end{array}
\right\}.
\end{eqnarray} 
We also note that for spheres the decrease of the \FWHM{} from the
maximum to that of the critical value is also stronger than in case of
the cloud radius. The value of a critical stable sphere is given by:
\begin{equation}
	\FWHMEQ_{\rm sph}^{\rm crit} =0.106\left(\frac{T}{10~{\rm K}}\right)
		\left(\frac{p_{\rm ext}/k}{2\times 10^4~{\rm K~cm}^{-3}}\right)^{-\frac{1}{2}}~{\rm pc}.
\end{equation}

\subsubsection{The \FWHMEQ-$N_{\rm H}(0)$ relation}

\begin{figure}[htbp]
	\includegraphics[width=\hsize]{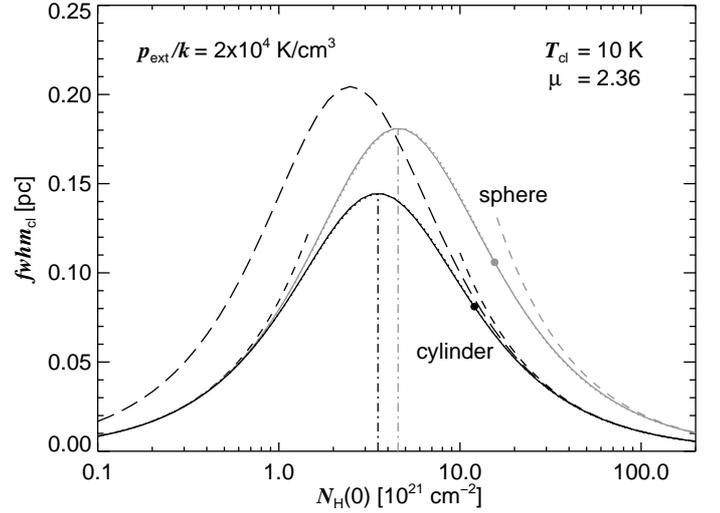}
	\caption{\label{fig_fwhmcolumn} Relation of the \FWHM{} and the
          central column density $N_{\rm H}(0)$ through an isothermal
          self-gravitating cylinder (black) and sphere (grey).  The
          cloud temperature is $10~{\rm K}$ and the external pressure
          is $2\times 10^4~{\rm K~cm}^{-3}$, except for the
          long-dashed curve where half the pressure is assumed.  The
          column densities corresponding to the maxima in the \FWHM{}
          are marked with vertical dashed-dotted lines.  The
          asymptotic behaviour for low and high overpressure is shown
          by short-dashed curves. The dotted curves (almost hidden by
          the solid curves) are approximations using
          Eq.~\ref{eq_fwhmcolumn}. The values for a cloud with the
          overpressure of a critical stable sphere are shown as filled
          circles. }
\end{figure}

There is a relationship between the two observables \FWHM{} and $N_{\rm
  H}(0)$, as shown in Fig.~\ref{fig_fwhmcolumn} for both spheres and
cylinders. $\bar N_{\rm H}(0)$, the column density corresponding to
the above $\FWHMEQ_{\rm max}$, is given by
\begin{eqnarray}
	\bar N_{\rm H}(0) &=& 10^{21}\left(\frac{p_{\rm ext}/k}{2\times 10^4~{\rm K~cm}^{-3}}\right)^{\frac{1}{2}}
		\nonumber \\
		&&\times  \left\{
\begin{array}{cc}
		3.524\,  {\rm cm^{-2}}, & f_{\rm cyl} = 0.331 \\[0.05cm]
		4.552\,  {\rm cm^{-2}}, & f_{\rm sph} = 0.521
\end{array}
\right\},
\end{eqnarray}
which is independent of $K$, and using Eq.~\ref{eq_fwhmmax} for cylinders
\begin{equation}
	\label{fwhmnh10max}
	\FWHMEQ^{\rm max}_{\rm cyl} \approx 0.051\, 
	\left(\frac{T}{10~{\rm K}}\right) \left(\frac{1 \times 10^{22} {\rm cm}^{-2}}{\bar N_{\rm H}(0)}\right)\, {\rm pc}.
\end{equation}

For high mass ratios ($f_{\rm cyl}>\sim 0.7$) if we eliminate the
dependence on $(1-f_{\rm cyl})$ in approximation
Eq.~\ref{eq_fwhmapprox} for the \FWHM{} using the inverse dependence in
approximation Eq.~\ref{eq_columnapprox} for the central column
density we obtain the simple relationship (see also
Eq.~\ref{fwhmandsigma1} in Appendix~\ref{sect_fwhmspheres})
\begin{equation}
	\label{fwhmnh1}
	\FWHMEQ_{\rm cyl} \approx 2 \sqrt{2^{2/3}-1}  (K/G) / (N_{\rm H}(0)\, \nrho),
\end{equation}
which is independent of both $p_{\rm ext}$ and $f_{\rm cyl}$ but does
respond to $K$.  For molecular gas, we have in the case of high
overpressure
\begin{equation}
	\label{fwhmnh10}
	\FWHMEQ_{\rm cyl} \approx 0.111 \, 
\left(\frac{T}{10~{\rm K}}\right) \left(\frac{1 \times 10^{22} {\rm
      cm}^{-2}}{N_{\rm H}(0)}\right)\, {\rm pc}.
\end{equation}
Note that this is identical to Equation~\ref{fwhmnh10max}, except for
a larger prefactor.

At small overpressure (flat density profile) the \FWHM{} and $N_{\rm
  H}(0)$ increase for given temperature and external pressure
proportionally with the cloud size. We obtain the asymptotic behaviour
for low mass ratio or low overpressure (see also
Equations~\ref{eq_columnapprox0}, \ref{eq_fwhmapprox0}, and
\ref{fwhmandsigma0})
\begin{equation}
	\label{fwhmnh0}
	\FWHMEQ_{\rm cyl}(f_{\rm cyl} \rightarrow 0) = \frac{\sqrt{3}}{2}
        \frac{K}{p_{\rm ext}}N_{\rm H}(0)\, \nrho.
\end{equation}

For intermediate values of the mass fraction, the  \FWHMEQ-$N_{\rm H}(0)$
relation for both of cylinders and spheres can be interpolated to high
accuracy between the above limiting behaviours, as
discussed in Appendix~\ref{sect_appC}.  The relation is
\begin{equation}
	\label{eq_fwhmcolumn}
	\FWHMEQ(t) = \FWHMEQ_{\rm max} C \frac{(t/t_0)^{a-1}}{(1+(t/t_0)^\gamma)^{(a+b)/\gamma}},
\end{equation}
where $t=N_{\rm H}(0)/\bar N_{H}(0)$ is the normalized central column
density, $a$, $b$, $t_0$, and $C$ are appropriate constants described
in the Appendix, and the power $\gamma$ is adjusted to provide the
best fit. The corresponding values for molecular cylinders and spheres
are given in Table~\ref{table_fwhmfit}, for $T=10~{\rm K}$ and the
adopted external pressure (denoted with superscript $0$).
For cylinders and spheres the analytical function is an excellent
approximation of the numerical solution overplotted in
Fig.~\ref{fig_fwhmcolumn}. For cylinders the approximation also
provides the correct power law behaviour for small and high
overpressure, by construction.  For the spheres we allowed for a
variation of index of the power law behaviour for large overpressure;
however, the change is small.

\begin{table}[htbp]
	\begin{tabular}{lccccccc}
		&$\FWHMEQ^0_{\rm max}$ & $\bar N^0_{\rm H}(0)$ & $\gamma$ & $a$ & $b$ & $C$\\
		&$[\rm pc]$ & $[10^{21}{\rm cm^{-2}}]$ & & & \\
		\hline
		cyl. &0.144  & 3.581 & 1.862 & 2 & 0.000 & 2.106\\
		sph. &0.181  & 4.648 & 1.796 & 2 & -0.044 & 2.126
	\end{tabular}
	\caption{\label{table_fwhmfit} Derived parameters of the approximation Eq.~\ref{eq_fwhmcolumn}
          of the \FWHMEQ-$N_{\rm H}(0)$ relation for spheres and cylinders. The temperature is $10~{\rm K}$
          and the external pressure $p_{\rm ext}/k = 2\times 10^4~{\rm K~cm^{-3}}$; 
          parameters for adopted values are denoted by superscript $0$.}
\end{table}

From Appendix~\ref{sect_appC} follows the scaling relation of the column
density at the maximum
\begin{equation}
	\frac{\bar N_{\rm H}(0)}{\bar N_{\rm H}^0(0)} = \left({\frac{p_{\rm ext}/k}{2\times 10^4~{\rm K~cm}^{-3}}}\right)^{\frac{1}{a+b}},
\end{equation}
and the scaling relation for the maximum \FWHM{}
\begin{equation}
	\frac{\FWHMEQ_{\rm max}}{\FWHMEQ^0_{\rm max}} = \left(\frac{T}{10~{\rm K}}\right) \left(\frac{p_{\rm ext}/k}{2\times 10^4~{\rm K~cm}^{-3}}\right)^{-\frac{b+1}{a+b}}.
\end{equation}
Combining these with the values in Table~\ref{table_fwhmfit} we
recover Equation~\ref{fwhmnh10max}.

For constant $p_{\rm ext}$, $\bar N_{\rm H}(0)$ is also fixed.  The
dependence of both asymptotes and $\FWHMEQ_{\rm max}$ on $K$ or $T$ is
the same and so the locus of \FWHMEQ\ vs.\ $N_{\rm H}(0)$ is simply
scaled by $K$.

\subsection{Dependence on external pressure}\label{sect_pressure}

\begin{figure*}
	\includegraphics[width=0.49\hsize]{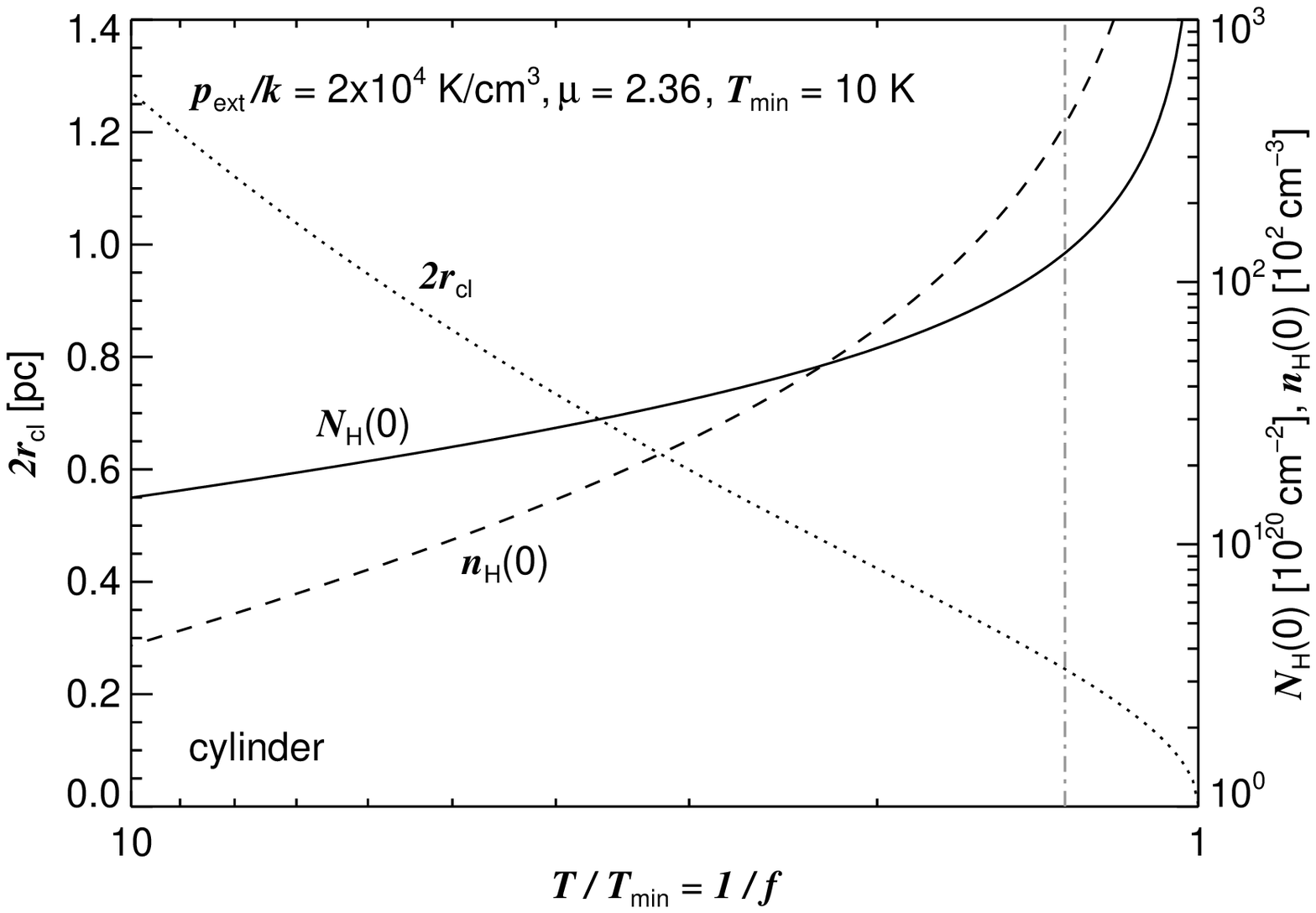}
	\hfill
	\includegraphics[width=0.49\hsize]{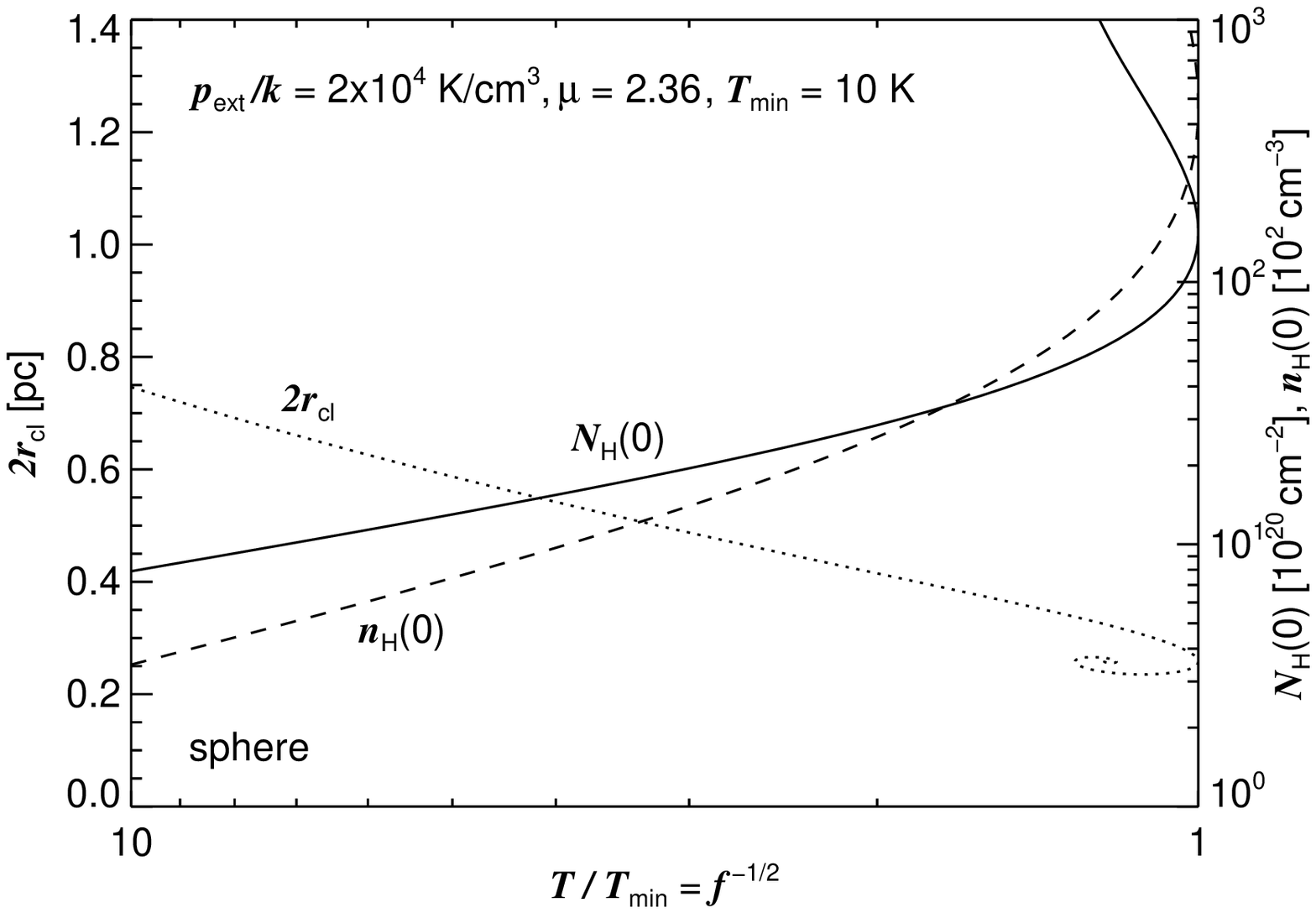}
	\caption{\label{figcooling} Physical parameters of an
          isothermal self-gravitating cylinder (left panel) and sphere
          (right panel) with fixed $p_{\rm ext}/k=2\times 10^4~{\rm
            K~cm}^{-3}$ as a function of cloud temperature.  Plotted
          are the cloud diameter ($2 r_{\rm cl}$, dotted line, scale
          on left axis), the central density (dashed line, right
          axis), and the column density through the cloud centre
          (solid line, right axis).  The values given for the size and
          the density depend on the minimum temperature $T_{\rm min}$
          assumed to be $10~{\rm K}$.  For spherical clouds the
          parameters are shown for all physical solutions including
          the supercritical ones characterized through an overpressure
          larger than $p_{\rm c}/p_{\rm ext}\approx 14.04$. The grey
          dashed-dotted line in the left hand figure gives the
          temperature ratio for $f_{\rm cyl}=0.75$.}
\end{figure*}

In the previous sections we have examined how the physical properties
of isothermal self-gravitating cylinders and spheres, subject to a
fixed external pressure and with a fixed $K$ or temperature, change
with the mass, parameterized through the mass ratio $f$.  We can
therefore directly understand how the properties would be affected by
a change of mass, say through accretion of additional material.

In this and the following subsection we discuss how the physical
parameters are affected by altering the external pressure and the
temperature (through cooling), respectively.

\subsubsection{Changing external pressure}\label{compression}

As we have seen in Sect.~\ref{massline}, the maximum mass line density
of a cylinder does not depend on the external pressure. Therefore, in
the case of constant mass line density $f_{\rm cyl}$ and fixed
temperature, a change of the external pressure has no effect on the
gravitational state.  In a quasi-equilibrium situation a cylinder can
always compensate for a higher external pressure through compression;
the central pressure rises so that the overpressure
(Eq.~\ref{eq_centralpressure}) remains the same.  While a change of
the external pressure has no effect on the density profile it does
affect the appearance because a cylinder becomes more opaque in higher
pressure regions in accordance with $N_{\rm H}(0) \propto \sqrt{p_{\rm ext}}$ 
(Eq.~\ref{eq:nh0}).

The case of a sphere with constant temperature and constant mass $M$
is quite different; a change of the external pressure affects the
gravitational state.  The ratio of $M$ to the critical mass changes as
$f_{\rm sph} = \sqrt{p_{\rm ext}/p_{\rm ext}^{\rm crit}}$, where
$p_{\rm ext}^{\rm crit}$ is the maximum possible pressure that the
cloud of fixed $M$ can sustain at its edge (see Eq.~\ref{mbe}).
As the external pressure is increased the central pressure reaches a
finite maximum $14.04 p_{\rm crit}$ when $f_{\rm sph} = 1$,
after which there is no stable configuration.

%
Furthering the discussion in Sect.~\ref{massline}, a cloud is
considered to be stable if the compression leads to a pressure
increase at the cloud edge, i.e., the response $-({\rm d} p(r_{\rm cl})/{\rm d} r_{\rm cl})>0$ 
where $r_{\rm cl}$ is the cloud radius.  Subcritical spheres, those with overpressure
below 14.04, are stable against compression with a pressure response
$-({\rm d} p(r_{\rm sph})/{\rm d} r_{\rm shp}) >0$.  Just above the critical
state, a sphere is unstable.\footnote{An interesting consequence is
  that if in a molecular cloud a large fraction of gas were bound in
  already near-critical stable cores, a pressure enhancement caused by
  a supernova shock or stellar wind would lead to a sudden burst of
  star formation in the whole cloud.} However, we also note that
supercritical spheres significantly above the critical overpressure
are not all unstable according to the above definition.

In the case of a cylinder the response is
simply given by
\begin{equation}
	-\frac{{\rm d} p(r_{\rm cyl})}{{\rm d} r_{\rm cyl}}  = 2 \frac{p(r_{\rm cyl})}{r_{\rm cyl}} > 0
\end{equation}
for all mass ratios $0<f_{\rm cyl}<1$.  Because of this, all
gravitational states are stable.  In the context of the equilibrium
solutions being discussed, there is no critical state (or
supercritical states) for cylinders as there is for spheres (see also
\citealp{Ebert1955,McCrea1957}).

\subsubsection{Embedded filaments}\label{embedded}

Filaments are often characterized by comparing their mass line density
to the maximum value $2K/G$.  A basic problem of interpretation which
arises concerns the effective temperature of the filament, in
particular if the filament itself is not entirely isothermal.  For
example, larger filaments with a relatively high effective temperature
of 50~K or more might form/contain small dense interior filaments
which are a factor five or more colder.  The profile of the column
density would show a narrow distribution on top of a broader one.  The
small size of the narrow feature is related to the colder gas
temperature and to its higher external pressure, which in this case is
the pressure in the interior in the broad filament.  Depending on the
distribution of mass assigned, both broad and narrow filaments might
be below their relevant maximum mass line density (the fact that they
were observed would suggest this conclusion).  See further discussion
in Sects.~\ref{largef} and \ref{highmass}.

\subsection{Dependence on temperature}\label{sect_temperature}

To understand how the isothermal self-gravitating clouds are affected
though cloud cooling we consider a spherical cloud of fixed mass and a
cylindrical filament with fixed mass line density. For a given
external pressure, equilibrium solutions of isothermal
self-gravitating clouds exist only above a temperature $T_{\rm min}$.
For a sphere, $T_{\rm min}$ is the temperature corresponding to the
critical stable cloud, from Eq.~\ref{mbe} for the fixed mass.  For a
cylinder, $T_{\rm min}$ derives from Eq.~\ref{eq_maxmasslinedensity}
for the fixed mass line density.  Mathematically we can express ratio
of the temperature to $T_{\rm min}$ in terms of $f$ and so the
gravitational state of the cloud is given by $T/T_{\rm min}\ge 1$.
For spheres $T/T_{\rm min} = 1/\sqrt{f_{\rm sph}}$ and for cylinders
$T/T_{\rm min}=1/f_{\rm cyl}$.

The cloud parameters (diameter, central density, and column density)
for both cylinders and spheres are shown in Fig.~\ref{figcooling}
where as a minimum temperature we have chosen $T_{\rm min}=10~{\rm K}$.
As the cloud cools at fixed mass line density and $p_{\rm ext}$, its
size shrinks while both central density and the column density through
the centre rise.  The variation of the radius and central density with
temperature is stronger in the case of a cylinder compared to a sphere.
For a given temperature ratio, the central column density through a
cylinder is larger than that through a sphere if we neglect the
complication of supercritical stable cases.  

As it cooled to $T_{\rm min}$, a cylinder would approach
asymptotically a singular state of an infinitely thin cylinder with
infinite overpressure, assuming it was not in the meantime fragmented
through instabilities.  At a lower temperature still, there would not
be an equilibrium solution at all: it would collapse to a spindle.

Through cooling a sphere reaches the critical point beyond which
further cooling would cause a pressure drop at the cloud edge so that
the cloud would collapse in a free fall time.  The physical solutions
presented in Fig.~\ref{figcooling} for pressurized supercritical
spheres correspond to cloud temperatures above the critical
temperature; these supercritical spheres are not accessible simply 
by cooling at fixed mass and external pressure.

For fixed low mass or low mass line density $T_{\rm min}$ might lie at
a value that cannot be reached by any efficient cooling mechanism.
For example Bok globules show temperatures not much lower than typical
10~K \citep{MyersBenson1983,BensonMyers1989}.  See also
Appendix~\ref{scaletemperature}.
Clouds with $T_{\rm min}$ below this empirical physical limit
would therefore remain as gravitational stable equilibrium
configurations.  A filament might possibly enhance its mass line
density through accretion, raising $T_{\rm min}$ and obviating this
cooling barrier.  Such spheres would also need to accrete more
material to become physically unstable and/or they might become
unstable through increased compression which raises $T_{\rm
    min}$ as $p_{\rm ext}^{1/4}$.

\section{\label{Sect_observations}Observed physical parameters of filaments}

In the following we validate the model by comparing the theoretical
prediction with recent observations.
This is important to confirm because then application of this model to
interstellar filaments would allow an independent estimate of
astrophysical quantities.  For example, it could be used to estimate
the interstellar pressure and the distance to cloud complexes and star
formation regions, and it could be used in combination of measurements
of the dust emission to estimate the dust emissivity.

\begin{figure}[htbp]
	\includegraphics[width=\hsize]{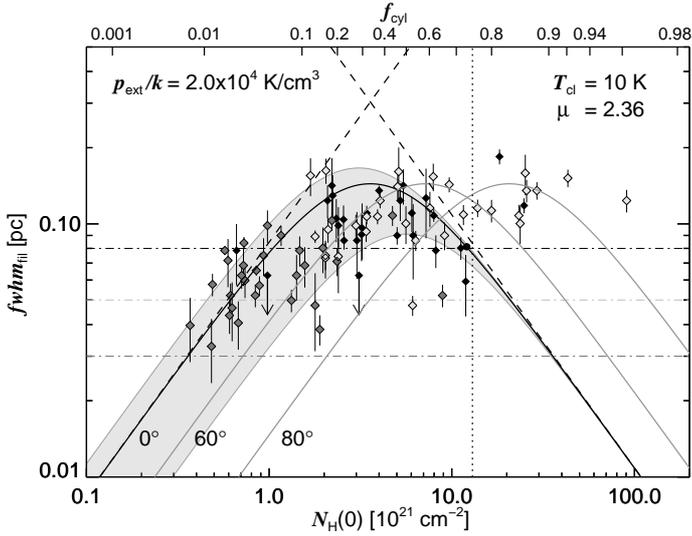}
	\caption{\label{fig_columnfwhmobserv} \FWHMEQ-$N_{\rm H}(0)$
          relation of filaments.  Observed values in \object{Polaris}
          (dark grey symbols), \object{IC 5146} (black symbols), and
          \object{Aquila} (light grey symbols), from Fig.~7 of
          \citet{Arzoumanian2011}. Note that we are plotting the
          column density of H nucleons rather than H$_2$.  We also
          lowered their values by 2.33/2.8 to correct for the value of
          $\bar \mu$ that they adopted. The horizontal dashed-dotted
          lines give the imaging resolution achieved in the three
          fields (black: \object{IC 5146}, dark grey: \object{Polaris},
          light grey: \object{Aquila}).  Predictions from the model of
          pressure-confined isothermal self-gravitating cylinders are
          given for three different inclination angles ($0^\circ$,
          $60^\circ$, and $80^\circ$; solid black and grey curves).
          The cloud temperature of the molecular gas is taken to be
          $T=10~{\rm K}$.  The external pressure provided by the
          ambient medium is assumed to be $p_{\rm ext}/k = 2 \times
          10^4~{\rm K~cm}^{-3}$ (see discussion in
          Appendix~\ref{scalepressure}).  The dashed lines are the
          asymptotes for low and high overpressure $p_{\rm c}/p_{\rm
            ext}$ or $f_{\rm cyl}$ (this is a logarithmic version of
          Fig.~\ref{fig_fwhmcolumn}).  For an overpressure
          corresponding to that of a critical stable sphere, and for
          inclination $0^\circ$, the expected value is shown as the
          filled black circle.  Likewise, the vertical dotted line
          indicates the column density of a cylinder with mass ratio
          $f_{\rm cyl}=0.75$; the upper non-linear scale indicates
          other values of $f_{\rm cyl}$ for this inclination. The
          expected range for a pressure regime 1.5 to $5 \times
          10^4~{\rm K~cm}^{-3}$ is shown by light grey shading.  }
\end{figure}

Imaging submillimetre observations with \emph{Herschel} provide
estimates of the \FWHM{} of the filaments, their density profiles, and
of the central column densities.  Some uncertainties relating to such
measurements are discussed in Appendix~\ref{App_uncertainty}.
Here, we discuss the results for individual filaments in \object{Polaris},
\object{IC 5146}, and \object{Aquila} presented by \citet{Arzoumanian2011}.  To
describe the filament profiles, they adopted a generalization of
Eq.~\ref{eq_densityprofile} for the density profile, parameterized by
index $\pindex$:
\begin{equation}
	\rho(r) = \rho_{\rm c} (1+(r/\sqrt{8}r_0)^2)^{-\pindex/2}.
\end{equation}
At large cloud radii ($r\gg \sqrt{8}r_0$) the profiles of the
filaments in \object{IC 5146} indicated $\pindex \sim 1.6\pm 0.3$, less steep
than the value four they expected for clouds in a vacuum, in agreement
with previous measurements discussed by \citet{FiegePudritz2000a}.
Our interpretation is simply that an equilibrium cylinder that is
pressure confined is the relevant model.  Relatively flat profiles are
a natural outcome for cylinders with a mass ratio $f_{\rm cyl}$
considerably smaller than unity (Section~\ref{radialdensity}).

In Fig.~\ref{fig_columnfwhmobserv} the deconvolved \FWHM{}
and $N_{\rm H}(0)$ results of \citet{Arzoumanian2011} are compared
with the self-consistent prediction of our model of pressure-confined
cylinders, including the asymptotic behaviours.
The parameters for the filaments of the three observed regions occupy
different areas in the \FWHMEQ-$N_{\rm H}(0)$~plane.
The filaments in \object{Polaris} show the lowest central column densities
with most values below $N_{\rm H}(0)< 10^{21}~{\rm cm^{-2}}$. The
sizes are also apparently smaller compared to the two other
samples. The filaments in \object{IC 5146} have intermediate central column
densities with a mean $N_{\rm H} \sim 6\times 10^{21}~{\rm cm^{2}}$.
The highest column densities were observed for filamentary structures 
in \object{Aquila} with values up to $ \sim 10^{23}~{\rm cm^{-2}}$.
Overall, the data indicate a trend that filaments with low column
densities are on average smaller than those with high column densities. 

As we see in Fig.~\ref{fig_columnfwhmobserv} the model of pressurized
cylinders shows a good agreement with the observations, although
obvious deviations do exist.  The observations are broadly consistent
with a cloud temperature $T=10~{\rm K}$ and an ambient pressure of
$p_{\rm ext}/k = 2\times 10^4~{\rm K~cm}^{-3}$.  Such a pressure is in
the range expected, as discussed in Appendix~\ref{scalepressure}.  The
data show a dispersion around the theoretical curve which can be
attributed to a variation of the external pressure alone.  Most data
points lie in a pressure regime $p_{\rm ext}/k$ from $1.5\times
10^4~{\rm K~cm}^{-3}$ to $5\times 10^4~{\rm K~cm}^{-3}$.  Note how the
position of the maximum lies on a line parallel to the high $f_{\rm
  cyl}$ asymptote.  An additional horizontal deviation of data from
the model is expected because of the distribution of inclination
angles; as shown in Fig.~\ref{fig_columnfwhmobserv}, higher
inclination could in principle explain some of the high values of the
column densities.  Note that at constant pressure, the \FWHMEQ-$N_{\rm
  H}(0)$ curve translates vertically with $T$.

\citet{Kandori2005} studied the physical parameters of Bok Globules
using an idealized model of a pressurized isothermal self-gravitating
sphere to fit the column density profile obtained from extinction
measurements.  The derived external pressures ranged from $2.1\times
10^4~{\rm K~cm}^{-3}$ to $1.8\times 10^5~{\rm K~cm}^{-3}$ with a mean
of $5.7\times 10^4~{\rm K~cm}^{-3}$.  There is overlap with the
pressures derived here but on average they are higher.

\subsection{Behaviour at small column density}\label{smallf}

Comparison of the data to the asymptotic behaviour at low $f_{\rm cyl}$
(Eq.~\ref{fwhmandsigma0}) constrains the ratio $T/p_{\rm ext}$.
In combination with this, the position of the maximum
(Eq.~\ref{fwhmnh10max}) constrains $T$.  At low column density, a
higher pressure could be accommodated by raising $T$, but for higher
column densities near the predicted maximum the \FWHM{} might become
too large.  Likewise, if no efficient cooling mechanisms exist to cool
the cloud temperature below $10~{\rm K}$, then we have a lower limit
on the required external pressure close to the cited values.

The parameters of most of the filaments in \object{Polaris} seem to follow
the relation expected for rather low mass ratio, $f_{\rm cyl}<0.1$.
Those structures are therefore not strongly self-gravitating and so
possibly transient density enhancements.  If they exist, filaments
with even lower $f_{\rm cyl}$ would be small compared to the
instrumental resolution and of low contrast in the images.
Stronger background fluctuations (cirrus noise; e.g.,
\citealp{Martin2010}) in the other two cloud complexes would produce a
more challenging limitation to identifying faint low column density
structures, and so the fact that filaments with low mass ratio are
less frequent or entirely absent in the two other samples might result
from this selection effect.

\subsection{Behaviour at large column density}\label{largef}

We think that inclination effects are unlikely responsible for the
extremely high column densities reported.  High column densities
without the expected small sizes might instead be related to cold
filaments which are embedded and pressure-confined within larger
filamentary structures, as mentioned in Sect.~\ref{embedded}.  Our
simple idealized model is therefore not easily applicable.

For example, in the \object{IC 5146} sample, the filament with the
highest column density (numbered 6 by \citealp{Arzoumanian2011}) also
has the highest reported mass line density (152~M$_\odot$~pc$^{-1}$),
\emph{much} higher than the expected maximum value for a filament with
$K$ corresponding to 10~K.  This is a very interesting but complex
filament with large variations in central column density and \FWHM{}
along the sinuous ridge.  The images and average line profile show one
or more cold irregular inner filaments inside a broader structure.
The narrow filaments are often paired in parallel segments.  How high
the mass line density is judged to be depends on how far out radially
the column density is integrated and this depends on where the adopted
model profile meets the ``background.''  Measurement of the embedded
filament(s) above the complex background of the broad filament would
produce physical parameters closer to the simple model.  A slightly
higher $K$ would help too, moving the predicted curve vertically in
Fig.~\ref{fig_columnfwhmobserv}.

Our sense is that this whole filament is not in free fall radial
collapse, in which case the self-gravity of the larger embedding
structure needs to be balanced too; an isothermal model is probably
not appropriate, but a higher $K$ (or perhaps even extra magnetic
pressure) would be needed.
This is open to experimental investigation by molecular velocity and
line-width measurements such as reported by \citet{Pineda2010}.
\citet{Kramer1999} have in fact mapped a 0.4~pc square region on the
ridge of this filament in some lines of the rare isotopologues of CO.
Even within this small region they find many clumps and two ridges
with differing velocites and line widths.  Overall this dense central
region, while clumpy, is fairly cold in terms of $K$ though perhaps
warmer than 10~K.  On the other hand, for $^{13}$CO on a larger scale
(2.5~pc) the velocity dispersion $\sigma \approx 1$~km~s$^{-1}$
(position C3 in \citealp{Dobashi1992}).  While perhaps exaggerated by
optical depth effects, this is suggestive of $K$ being an order of
magnitude larger in the broad embedding filament.  The situation seems
to be similar for the widest filament (numbered 12) where the profile
suggests an interior cold (sometimes paired) filamentary structure
surrounded by a more turbulent gas.

Interestingly, no filaments with high central extinction ($N_{\rm
  H}(0)>\sim 2\times 10^{22}~{\rm cm^{-2}}$) \emph{and} the
corresponding low size ($\FWHMEQ<0.08~{\rm pc}$) were found.  There is
no obvious selection effect against finding these high $f_{\rm cyl}$
filaments though, as suggested above, measurement of their parameters
might be an issue depending on their environment.

If strongly self-gravitating filaments, those above a certain mass
ratio, do not exist, then why?  This might be inherent to the process
responsible for the formation and evolution of filaments generally, or
might be because of their disruption and/or short lifetimes.  While
interesting, the former possibility is beyond the scope of this paper.
One clue to the latter possibility is that according
Fig.~\ref{fig_columnfwhmobserv} the empirical cut-off in $f_{\rm cyl}$
is near 0.75, where the cylinders have about the same overpressure as
a critical stable sphere and so if appropriately fragmented might form
structures that would collapse.  This is explored further in the
following section.

\section{Structure formation along filaments}\label{Sect_structure}

In most cases observed filaments show substructure, contributing to
the difficulty of measuring their parameters.  Most interesting for
understanding the potential role of filaments, as opposed to just high
column and volume density, in the star formation process are the
condensed structures extracted, ``prestellar cores'' having strong
self-gravity.  By definition they are cool, not having detectable
emission at 70~$\mu$m or 24~$\mu$m \citep{Konyves2010}, and so while
they have the \emph{potential} for forming protostars or young stellar
objects (YSOs) \citep{Andre2010} they are not yet vigorously doing so.
They are being externally heated (stage E, \citealp{roy2011}) with
little internal energy being generated.
These ``prestellar cores'' appear in the higher column density
filaments at typically $A_V > 8~{\rm mag}$, as in \object{Aquila} and
\object{IC 5146} \citep{Andre2010,Arzoumanian2011}.
Cold protostars are rarer and also often associated with filaments
\citep{Andre2010,Bontemps2010}.

Even filaments with much lower column density, those with low $f_{\rm
  cyl}$ such as in \object{Polaris}, have substructure.  These are dubbed
``starless cores'' as well, potentially confusing nomenclature:
although they have sizes typical of what defines cores (radii less
than a few tenths of a parsec) they comprise much less than a solar
mass and are not significantly gravitationally bound and so do not
seem destined to form stars \citep{Andre2010} .  

\subsection{Equilibrium filaments}

If the basic structure is described by an equilibrium filament, as we
have argued, then the question arises as to the origin of the
substructure.
Equilibrium filaments are subject to gravitational instabilities that
are axisymmetric, producing structure along the axial directions.  Two
types may be distinguished \citep{Nagasawa1987}.  For high $f_{\rm
  cyl}$ ($r_{\rm cyl} \gg \sqrt{8} r_0$) the instability forms out of
compression. A local density increase raises the self-gravity and
lowers the gravitational energy, leading to a growth in the amplitude
of the structure.  For low $f_{\rm cyl}$, those with a fairly flat
profile ($r_{\rm cyl} \ll \sqrt{8} r_0$), the instability relates to
deformation of the surface; in such a ``sausage instability'' it is
the volume energy of the surface integral of the external pressure
that is lowered.  Transitional behaviour mixing these cases occurs near
$f_{\rm cyl} = 1/2$ ($r_{\rm cyl} = \sqrt{8} r_0$).

\begin{figure}[htbp]
  \includegraphics[width=\hsize]{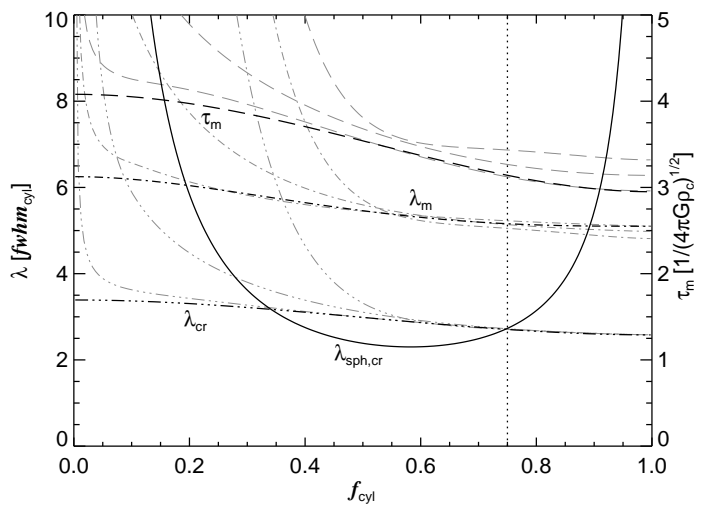}
  \caption{\label{fig_fragment} Length and time scales relevant to
    fragmentation as a function of $f_{\rm cyl}$, in units of the \FWHM{}
    (left axis) and $1/\sqrt{4 \pi G \rho_c}$, the sound crossing time
    of $r_0$ (right), respectively.  The shortest wavelength growing
    mode is $\lambda_{\rm cr}$ (dashed-multiple dotted lines).  The fastest growing mode is
    $\lambda_{\rm m}$ (dashed-single dotted lines) which, in the linear regime,  develops on 
    a time scale $\tau_{\rm m}$ (long-dashed lines).  For comparison, $\lambda_{\rm cr, sph}$ is
    the length along the filament required to encompass a mass equal
    to that of a critical stable sphere for the same $p_{\rm ext}$ and
    $K$ (solid line). The black broken lines correspond to non-magnetized
    cylinders. The related grey lines show the increasing effect of a
    magnetic field and correspond to $p^2=1\%$, $10\%$, and $100\%$ of
    equipartition.  The values are based on the work of
    \citet{Nagasawa1987}, see App.~\ref{Sect_polapprox}.  The vertical
    dotted line marks the mass ratio $f_{\rm cyl}=0.75$. }
\end{figure}

\begin{figure}[htbp]
  \includegraphics[width=\hsize]{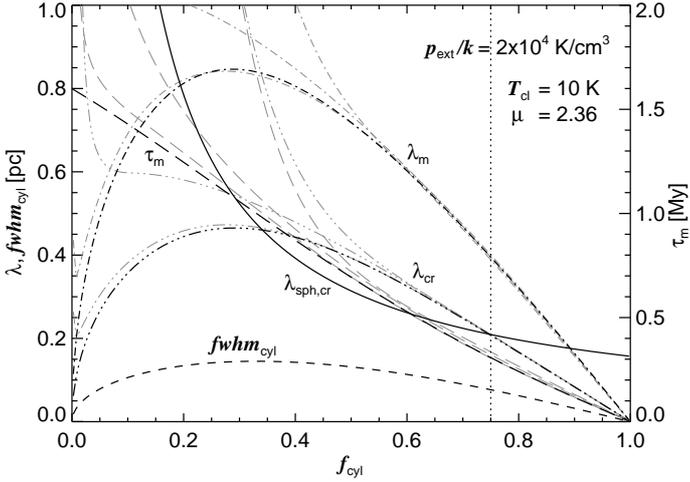}
  \caption{\label{fig_fragment2} Like
    Fig.~\ref{fig_fragment}, but with lengths and time scales
    expressed in physical units. The molecular cylinders are assumed 
    to have temperatures of $T_{\rm cl}=10~{\rm K}$ and  to be
    pressurized by a medium with $p_{\rm ext}/k=2\times 10^4~{\rm K/cm^3}$. 
    Also shown is the corresponding 
    \FWHM{} (see Fig.~\ref{fig_fwhm}).}
\end{figure}

Linear analysis of the dispersion relation \citep{Nagasawa1987} shows
that the critical wavelength above which perturbations grow is, for
$f_{\rm cyl} \rightarrow 1$,
\begin{equation}
	\label{vacuum}
	\lambda_{\rm cr} = 3.96\, \sqrt 8 r_0 = 3.96\, r_{\rm cyl} \sqrt{(1 - f_{\rm cyl})} = 2.58\, \FWHMEQ_{\rm cyl}
\end{equation}
and, for $f_{\rm cyl} \rightarrow 0$,
\begin{equation}
	\label{incompressible}
	\lambda_{\rm cr}= 5.87\, r_{\rm cyl} = 5.87\, \sqrt{f_{\rm cyl}} \sqrt{8} r_0 =3.39\, \FWHMEQ_{\rm cyl},
\end{equation}
with a smooth variation for values of $f_{\rm cyl}$ in between.  The
wavelength corresponding to the maximum growth rate, $\lambda_{\rm
  m}$, is about twice $\lambda_{\rm cr}$ (a factor 1.96 and 1.84 for
the above limiting cases, respectively).  Figure~\ref{fig_fragment}
shows the $f_{\rm cyl} $ dependence of $\lambda_{\rm cr}$ and
$\lambda_{\rm m}$, relative to the \FWHM.  At least initially, the
fastest growing mode is quite elongated, with $\lambda_{\rm m}/\FWHMEQ_{\rm cyl}
\approx 5$.  This would also be the separation of the cores, in units
of \FWHM.  But in absolute terms, both $\lambda_{\rm m}$ and \FWHM{}
become arbitrarily small in the two limits.  This is shown in
Figure~\ref{fig_fragment2}.

An important issue is the growth time scale, which is proportional to
$1/\sqrt{4 \pi G \rho_{\rm c}} = r_0/\sqrt{K}$, the sound crossing
time across $r_0$.  As shown in Fig.~\ref{fig_fragment}, for the
fastest growing mode the constant of proportionality falls from 4.08
to 2.95 as $f_{\rm cyl}$ increases from 0 to 1 \citep{Nagasawa1987}.
It is also convenient to define a time scale $\tau_{\rm Kp} =
\sqrt{K/(4\pi G p_{\rm ext})}$, that evaluates to 0.4~My for the
adopted values of $K$ and $p_{\rm ext}$.  Thus for $f_{\rm cyl}
\rightarrow 1$,
\begin{equation}
\tau_{\rm m} = 2.95 (1-f_{\rm cyl}) \tau_{\rm Kp}.
\label{tau1}
\end{equation}
This clearly arbitrarily short.  For $f_{\rm cyl}$ near 0 the time
scale is
\begin{equation}
\tau_{\rm m} = 4.08 \tau_{\rm Kp}.
\label{tau0}
\end{equation}
While this is much faster than the crossing time $r_{\rm
  cyl}/\sqrt{K}$, because $r_{\rm cyl} \ll r_0$, it is still finite,
of order 1.6~My for the canonical values, even longer for a warmer or
more turbulent gas.  This would set up ``sausage instabilities.''  The
time scale $\tau_{\rm m}$ in My is shown in
Figure~\ref{fig_fragment2}.

The effect of a uniform magnetic field along the filament has also
been studied by \citet{Nagasawa1987}.  The ``compressive instability''
(finite $f_{\rm cyl}$) is not much affected because the field is
axial, but the ``sausage instability'' (low $f_{\rm cyl}$) is strongly
suppressed because the field limits the surface distortion.  In that
case, both $\lambda_{\rm m}$ (see Fig.~\ref{fig_fragment}) and the
time scale $\tau_{\rm m}$ are increased significantly, even for a
field much smaller than equipartition strength (where $p^2 = B^2/(4
\pi \rho_c K) = 1$).  Thus, in the ISM this instability seems unlikely
to be relevant to determining the substructure seen in filaments with
low mass fraction as in \object{Polaris}.

The question of the final outcome of the non-linear growth of the
``compressive instability'' is of course critical.  \cite{Curry2000a}
found a sequence of equilibrium structures within filaments.  Their
shape is prolate along the cylinder axis, within a tidal lobe within
which the equidensity contours are closed. These equilibria are
gravitationally dominated and so are rounder as the pressure (density)
contrast between centre and lobe increases (limiting axial ratio about
1.7).  The behaviour of the lobe radius and the mass contained within
the lobe with increasing pressure contrast is reminiscent of the
Bonnor-Ebert sequence, though it is not known whether these cores
become unstable to collapse at similar overpressures.

The non-linear growth of the instability has been followed numerically
by \cite{Inutsuka1997} for two cases of equilibrium cylinders with
$f_{\rm cyl} = 0.9$ and 0.2.  For the former they find a complete
disintegration of the filament into well separated spherical clouds
suggestive of gravitational collapse.  For the latter a distinctive
spherical core forms as well, with the central density increased by a
factor nine, but it does not collapse.

\subsubsection{Mass of the fragment}

The maximum mass available to be concentrated by a perturbation of
wavelength $\lambda$ is
\begin{equation}
	M_{frag} = \lambda M/l =  2 \sqrt{8} f_{\rm cyl} (1-f_{\rm cyl})
        \frac{\lambda}{\sqrt{8} r_0} \frac{K^2}{\sqrt{4\pi G^3 p_{\rm
              ext}}}.
\label{massfrag}
\end{equation}
This is deliberately cast in the same form as Eq.~\ref{mbe} for the
mass of a critical stable sphere, where the prefactor is simply 4.191.

Even if the growth of the perturbation is effective in concentrating
material into a fairly spherical core, it seems intuitive to us that
the outcome is unlikely to be gravitational collapse unless the mass
of the core is comparable to the critical Bonnor-Ebert mass for the
external pressure confining the original cylinder.  To accumulate the
critical mass, the required wavelength $\lambda_{\rm \cBE}$ is, when
$f_{\rm cyl}$ is close to 1,
\begin{equation}
	\lambda_{\rm \cBE} /\FWHMEQ_{\rm cyl} = 4.191 /\left(4 \sqrt{8} \sqrt{2^{2/3} -1} f_{\rm cyl}
	(1-f_{\rm cyl}) \right)
\label{lam1}
\end{equation}
and, when $f_{\rm cyl}$ is close to 0, 
\begin{equation}
	\lambda_{\rm \cBE} /\FWHMEQ_{\rm cyl}  = 4.191/\left(2 \sqrt{3} \sqrt{8} f_{\rm cyl}  \sqrt{f_{\rm cyl}  
	(1-f_{\rm cyl})} \right).
\label{lam0}
\end{equation}
Thus for both of these limiting cases the perturbations would
initially have to be extremely elongated compared to the \FWHM.  
The locus of $\lambda_{\rm \cBE} /\FWHMEQ_{\rm cyl}$ in Figure~\ref{fig_fragment}
contrasts these extreme values with the finite value of $\lambda_{\rm
m}/\FWHMEQ_{\rm cyl}$ (in the range 5 -- 6).  Thus in both limits of $f_{\rm
cyl}$, any condensations resulting from perturbations would be of low
mass compared to the critical Bonnor-Ebert mass; this is not a
promising scenario for forming prestellar or protostellar cores.

Indeed none are observed for filaments with low $f_{\rm cyl}$.  Also,
because of suppression by the magnetic field, we are less interested
in the case $f_{\rm cyl}$ close to 0.
It is hard to imagine a perfectly uniform filament being created in a
turbulent medium; whether the structure that does exist is just a
natural by-product of formation of the filament needs to be
investigated.

For filaments with very high overpressure, or $f_{\rm cyl}$ close to
unity, the implication is that a series of small dense condensations would
develop, each of insufficient mass to collapse.  On a longer time
scale these might merge to form more massive cores
\citep{Inutsuka1997}, perhaps arriving at the high density equilibrium
configuration found by \citet{Curry2000a}. 
According to the findings presented in Fig.~\ref{fig_columnfwhmobserv}
this $f_{\rm cyl} \rightarrow 1$ limiting case has little relevance
to the interstellar medium as the precursor narrow high column
density filaments have not been observed. How far the results are
biased through the method applied to characterize highly opaque filaments
need to investigate as well.

However, for \emph{intermediate} $f_{\rm cyl}$ we see in
Figure~\ref{fig_fragment} that $\lambda_{\rm \cBE}$ can actually be
less than $\lambda_{\rm m}$ (and even $\lambda_{\rm cr}$) and
comparable to the \FWHM.  This is a more propitious situation,
starting from a much less elongated perturbation, to accumulate a mass
comparable to the critical Bonnor-Ebert mass.

For such intermediate $f_{\rm cyl}$, $\tau_{\rm m} \approx 3.5
(1-f_{\rm cyl}) \tau_{\rm Kp}$.  This is sufficiently rapid
(Fig.~\ref{fig_fragment2}) that initially uniform filaments could
develop structure on astronomically relevant timescales.  As discussed
above, it seems plausible that some of these could accumulate a
critical Bonnor-Ebert mass and become the observed cold protostars.
If this were the explanation, then the protostars ought to be
separated by about $\lambda_{\rm m}$, about five times the \FWHM{} of
the embedding/undisturbed filament (Fig.~\ref{fig_fragment}).
This appears to be on the order of what is observed (see below) but
needs to be carefully quantified.

\subsection{Pearls on a string}\label{pearls}

According to \citet{Mensch2010}, the above-mentioned prestellar cores,
observed along high column density filaments, are about the same size
as the width of the filament.  Inspection of the images of source
positions presented \citep{Mensch2010,Arzoumanian2011} shows that they
are also separated by about the \FWHM{} of the filament.  Therefore,
remarkably, they are arranged ``\emph{almost like pearls on threads in
  a necklace}'' \citep{Mensch2010}.  Because of the small separation
it appears unlikely that those structures are caused directly by the
gravitational instabilities discussed above.  We have seen that
condensations resulting from that origin would have a separation of at
least 2.5 times the \FWHM{} (for intermediate mass ratios $f_{\rm cyl}
> \approx 0.5$).
We have examined the above-mentioned filament 6 in \object{IC 5146} in
both SCUBA \citep{diF2008} and \emph{Herschel} archival submillimetre
images.  There are many clumps that would be extracted as prestellar
cores, but not all are equal, and our impression is that these
are somewhat clustered with major concentrations typically separated
by 2' to 5', corresponding to 2.2 to 5.5 times the given \FWHM.  
  The YSOs associated with this filament found in the Spitzer study by
  \citet{Harvey2008} are also well separated.  This is perhaps more
in line with what would be expected from gravitational instabilities,
but then a hierarchy and perhaps time sequence of fragmentation is
suggested.

At this point, we feel that the origin of the substructure is an open
question.  Other than gravitational instability, perhaps it is a
natural consequence of the formation of such filaments; in the
presence of strong self-gravity, it might be difficult to form a
perfectly straight and uniform cylinder.  In our experience it is also
a challenge to extract and characterize ``sources'' found as closely
spaced inhomogeneities along a filament.

Because a separation of one \FWHM{} is less than $\lambda_{\rm
  cr,sph}$ (Fig.~\ref{fig_fragment}), these prestellar cores, while
strongly self-gravitating, would probably be stable against collapse,
consistent with
their being starless.  Whether these presently starless clouds will
form any stars in the future and therefore be relevant to the initial
mass function of the stars is also an interesting open question.

\subsection{Filaments with high mass line density}\label{highmass}

For a filamentary structure with more than the maximum mass line
density (Eq.~\ref{eq_masslinedensity}), there is no equilibrium
solution.  If they could be created, somehow, because of their high
density they would collapse on a rapid timescale.  

There are certainly filamentary configurations observed that have a
large mass line density, considerably greater than the maximum line
density \emph{if} evaluated for $K$ corresponding to a low value like
10~K which we find appropriate to simple narrow filaments.
\citet{Andre2010} conclude that the gravitational fragmentation of
such ``supercritical''\footnote{Despite our comments in
  Sect.~\ref{massline}, their terminology is adopted here in this
  subsection.} filaments is responsible for the formation of the
observed self-gravitating prestellar cores and protostars.  There are
two problems with this scenario.

First, our interpretation of the numerical calculations of
\citet{Inutsuka1997} is that there is fragmentation only if the mass
line density is finely tuned to being just above the maximum value
and/or if the initial conditions already contain large perturbations,
which begs the question.  Perhaps a relevant configuration could be
set up beginning with an equilibrium filament with $f_{\rm cyl} $
close to 1 and then making it ``supercritical'' by lowering the
temperature.  Otherwise, the predominant outcome is radial collapse of
the cylinder to a spindle.  Eventually the collapsing spindle would
become optically thick and could develop substructure; but this would
be on a small scale because of the large density and small size of the
collapsed filament \citep{Inutsuka1997}.

Second, in the process of forming the fragments, a ``supercritical''
filament itself would collapse, be consumed, and disappear on the same
rapid time scale.  However, the observations indicate that prestellar
cores, and even YSOs, occur along identifiable filaments of finite
size, even if it appears that the mass line density is
``supercritical.''  This coexistence is a challenge to be satisfied by
any model of the structure formation, even for initially equilibrium
filaments.

While high column and volume density are arguably fundamental to any
process of fragmentation, our conclusion is that protostar and cluster
formation must be a more complex process than simply the gravitational
fragmentation of ``supercritical'' filaments.  Based on the complex
radial profiles of these high mass line density filaments, we think
that some progress might be made by considering cold embedded
filaments within a broader structure supported by a gas with higher
effective $K$, in which case the overall configuration is not actually
``supercritical'' and in free fall.

\section{Summary and discussion}\label{Sect_summary}

We have analyzed the physical properties of interstellar filaments on
the basis of an idealized model of isothermal self-gravitating
infinitely long cylinders which are pressurized by the ambient
medium. The pressure-confined cylinders have a mass-line density
smaller than cylinders extending into a vacuum.  The mass fraction
$f_{\rm cyl}$, the ratio of the mass-line density to the maximum
possible for equilibrium configurations, describes the gravitational
state of the cylinder.  For given temperature and external pressure,
it is used in deriving analytical expressions for the central density,
the radius, the column density profile, the central and average column
densities and the \FWHM.  The dependence of the physical properties
on external pressure and temperature is clear in these expressions.
The results are compared to the case of pressure-confined isothermal
self-gravitating (Bonnor-Ebert) spheres, characterized by the mass
fraction $f_{\rm sph}$, the ratio of the mass to the mass of the
critical stable sphere.

We compared the model prediction for the relation between the size and
the central column density with recent observations of filaments.
Given the complexity seen it is gratifying to find good agreement,
even surprising considering the idealization of the model.  In
practice the model seems best applicable to those filaments that are
cold and show a simple smooth elongated structure.  On the basis of
the model the filaments appear to experience an ambient pressure
$p_{\rm ext}/k$ in the range $1.5\times 10^{4}~{\rm K~cm^{-3}}$ to
$5\times 10^4~{\rm K~cm}^{-3}$ and to have a gas temperature $T\approx
10~{\rm K}$.  For these parameters, the observations indicate an
apparent physical upper limit to the mass ratio $f_{\rm cyl} \approx
0.75$; higher column density narrower filaments are not seen.  The
corresponding maximum central extinction for the observed filaments
fitting the model is $A_V \approx 7~{\rm mag}$.  This is interestingly
close to the empirical threshold for star formation, but we think this
is coincidental, especially given the variable threshold from region
to region \citep{Enoch2007}.

We have summarized and discussed previous stability studies that have
shown that all unmagnetized infinitely long cylinders $f_{\rm cyl}<1$
are subject to gravitational instabilities.  While a magnetic field
can have a stabilizing effect for $f_{\rm cyl}<0.5$, in particular
suppressing ``sausage instabilities'' for $f_{\rm cyl}\approx 0$,
there is a negligible effect on compressive instabilities for $f_{\rm
  cyl}>0.5$.  Interstellar magnetized filaments would seem to become
increasingly unstable towards large $f_{\rm cyl}$ because of a
strongly increasing growth rate of the instabilities.  The
disintegration of filaments close to the maximum mass line density
($f_{\rm cyl}>0.9$) would produce fragments whose masses are well
below the critical value to form stars.  We found no indication in the
theory for a strong threshold for the formation of fragments that
would collapse.  However, intermediate values of $f_{\rm cyl}$ seem
most favorable.  The fastest growing disturbance for $f_{\rm cyl}>0.5$
occurs at a length approximately five times the \FWHM, so that
protostars or clusters would be well separated.

It is difficult to understand in this model why in high column density
filaments there would be substructure akin to the extracted prestellar
cores separated along the filament by only one \FWHM, which is well
below the minimum length at which a disturbance would grow
($\lambda_{\rm crit} >\approx 2.5\times \FWHMEQ_{\rm cyl}$).  This might point to
external influences accompanying the process that formed the filament
in the first place, resulting in the non-uniform mass line density
along these filaments, not the idealized model considered here.  On
the other hand, if one concentrates on the major mass concentrations
along the filament, their separation might be more compatible with an
origin in gravitational instability along the cylinder.

High mass line density filaments are observed that would appear to
have more than the maximum mass line density that could exist in
equilibrium, at least \emph{if} evaluated for the low adopted $K$.
The coexistence of embedded narrow high column density filaments, even
in pairs, and considerable substructure (prestellar cores) and
protostars suggests to us a scenario more complex than simply the
free-fall collapse and fragmentation of cylinders with more than the
maximum line density.  Although the simple model cannot explain such
complex structure, it points to a situation in which the embedding
structure is supported by gas with a $K$ larger than the low value
that might reasonably be adopted for the narrow embedded structure.
This is obviously open to observational scrutiny via mapping of
velocities and line widths of appropriate molecules.

As \citet{McCrea1957} has emphasized, ``genuine'' gravitational
collapse to form stars and clusters is best realized in configurations
with roughly the same dimensions in all directions so that
gravitational effects are three-dimensional.  He argued that the
breakup of less favorable one-dimensional configurations like
filaments must be due to irregularities in the density distribution
and the external pressure to which they are subjected, or to
differential motions.  Empirically, there is suggestive evidence that
such breaking of the symmetry of an idealized long cylinder is
important, as occurs for example where filaments appear to cross.
These are matters that could be explored now in the results of
numerical simulations.

\acknowledgement{ JF is thankful for financial support from CITA and
  the MSO.  Personally he likes to thank Prof. B. Schmidt and Prof. M. Dopita.
  This work was supported by grants from the Natural
  Sciences and Engineering Research Council of Canada and the Canadian
  Space Agency.}

\bibliographystyle{aa} 
\bibliography{fischerareference,additional}

\appendix

\section{\label{App_uncertainty}Questions of scale}

In comparing the models with the data we need to be aware of the
context, what can be measured and the related uncertainties.  In this
appendix we discuss column density, size, external pressure, and
effective temperature.

\subsection{Column density}\label{scalecolumn}

Column densities of filaments from \emph{Herschel} data are
fundamentally in units of the dust optical depth in the submillimetre.
This is converted to $N_{\rm H}$ using a submillimetre dust opacity.
This opacity appears to show a dependence on environment and might be
uncertain by a factor two \citep{Martin2012}.  For the overview
comparison in Figure~\ref{fig_columnfwhmobserv}, where $N_{\rm H}(0)$
is on a logarithmic scale, this is might not be a major concern.

In the diffuse and largely atomic interstellar medium, where both
$N_{\rm H}$ and optical interstellar extinction curves can be observed
directly, we have the ratio $N_{\rm H}/E(B-V)=5.8\times 10^{21}{\rm
  cm^{-2}/mag}$ \citep{Bohlin1978} and an absolute to relative
extinction $R_V= A_V/E(B-V) = 3.1$ \citep{Fitzpatrick1999}.  A column
density of $10^{22}~{\rm cm^{-2}}$ therefore corresponds to $A_V =
5.3$~mag.  We use this same conversion when $A_V$ is being used a
``shorthand'' for column density in dense molecular environments,
recognizing that the actual visual extinction, if it could be
measured, would probably not be this value.

Another measure more applicable at higher column densities is near
infrared colour excess.  This is also often converted to $A_V$, which again
should be regarded as a ``shorthand'' because the shape of the
interstellar extinction curve through the optical probably varies.  It
might be more reliable to convert the near infrared colour excess
directly to $N_{\rm H}$, but that calibration is not actually known
for the high column densities encountered in star-forming regions
\citep{Martin2012}.

Another uncertainty might be related to radiative transfer which
produces a radial variation of the dust temperature within
externally-heated filaments.  While the dust
opacity seems to be larger for dust in denser environments, the
averaging of the dust emission over a distribution of temperatures
leads to a reduced effective emission coefficient and a spectral
dependence of the emissivity that is less steep than is intrinsic for
dust in a diffuse environment as theoretically demonstrated for condensed
cores \citep{Fischera2011}. The radiative transfer effects increase
with central column density and could therefore play a larger role at
high mass ratios.  The column density profile indicated by the surface
brightness would be flatter than the intrinsic profile.

\subsection{Size}\label{scalesize}

Sizes of filaments are measured from the profile of the column density
in the direction orthogonal to the axis of the filament.  The \FWHM{}
could be measured directly but is more usually found as a parameter of
a model fit to the profile.  The model fit also needs to account for
the background on which the filament appears superimposed.  In
practice both the background and the profile can vary along the
filament, adding to the challenges (\citealp{Arzoumanian2011} resorted
to using a Gaussian model to find the \FWHM) and the reported standard
deviation in the parameters.  Even in nearby molecular clouds, the
filaments, while resolved, are often not a lot larger than the imaging
resolution.  Therefore, the reported sizes have been ``deconvolved.''
In the future, this might be avoided by forward modeling, adopting
analytical models like we have presented, convolving the predictions
to the resolution of the data, and then iteratively fitting the
parameters.

The background is still something to be fit too.  Depending on the
circumstances, this background could be from foreground or background
material unrelated to the filament, and/or a contribution from the
medium providing the pressure confinement.  A well-defined edge of a
filament might suggest, for example, that the filament can be regarded
as cold material which is surrounded by a warmer or more turbulent but
less dense environment.  This relates to the following discussion of
external pressure.

\subsection{External pressure}\label{scalepressure}

\citet{Boulares1990} presented a model of the vertical support of the
Galactic disk.  The mid-plane pressure near the sun is at least
$p_{\rm ext}/k=2.5\times 10^4$~K~cm$^{-3}$.  About a third of this is
the effective vertical support by cosmic rays; however, on the size
scale of the structures of interest here, cosmic ray pressure can be
ignored \citet{Curry2000,Fischera2008,Fischera2011}.  The remainder is
a combination of kinetic pressure (thermal and turbulent motions) and
magnetic pressure.  For a magnetic field of less than equipartition
strength, the kinetic pressure would be at least $0.8 \times 10^4~{\rm
  K~cm}^{-3}$.  The thermal pressure is only about a third of this
\citep{Kalberla2009}, the rest being turbulent pressure.

The filaments being discussed are within clouds with sufficient column
density to have become molecular ($ \left<A_V \right> > 1$; e.g.,
\citealp{Glover2011}).  If it is assumed that these clouds can
themselves be described as pressure-confined isothermal spheres, with
enough self-gravity to be near the critical mass, then following
Eq.~\ref{eq_meancolumn} and Eq.~\ref{eq_meansigma} the external
pressure required to produce a mass surface density
$\left<\Sigma\right>$ averaged over the projected radius is given by
\footnote{Using a dimensional analysis of a cloud in virial
  equilibrium, \citet{McKee2003} derive a similar relationship between
  the average \emph{internal} pressure and the column density.  There
  is a similar coefficient of proportionality, not surprisingly because
  both configurations are strongly self-gravitating and because the
  average pressure in a critical sphere is only 2.5 times larger than
  the external pressure \citep{Spitzer1968}.}
\begin{eqnarray}
	\label{eq_meanAv}
p_{\rm ext} & = & (\pi/8) G \left<\Sigma\right>^2 \, {\rm or}
\nonumber \\
p_{\rm ext}/k & = & 0.37 \times 10^4 \left<A_V \right>^2 ~{\rm K~cm}^{-3}.
\end{eqnarray}
A subcritical sphere would require a greater external pressure to
produce the same average column density
(Fig.~\ref{fig_columndensity}).  The average column density in the GMC
sample discussed by \citet{Heyer2009} corresponds to $\left< A_V
\right> \approx 2.8$, which would require a pressure of $2.9 \times
10^4~{\rm K~cm}^{-3}$.  This seems consistent with the fact that this
cloud sample is in the inner Galaxy where the midplane pressure is
probably higher.

The central extinction through a critical stable isothermal sphere is
$A_V = 8.0 \sqrt{(p_{\rm ext}/k)/(2\times 10^4~{\rm K~cm}^{-3}})$ mag
(\citealp{Fischera2008}; Fig.~\ref{fig_columndensity}).  This
predicted amount of extinction seems to be in agreement with that
observed for molecular clouds \citep{Dopita2003}.

Two conclusions are that the midplane pressure is indeed quite high,
and that the pressure within the molecular cloud due to its self
gravity -- the ambient external pressure experienced by the filaments
-- will be higher still, depending on where the filaments lie within
the pressure profile of the molecular cloud.

Another contribution could be due to ram pressure, for example related
to the sweeping up of the North Celestial Loop of which the
\object{Polaris} field is a part.   

\subsection{Effective temperature}\label{scaletemperature}

For the model we need the effective temperature embodied in $K$,
related to both the thermal and the turbulent (non-thermal) motion of
the gas.  The most direct way to measure this would be through the
line width of an appropriate molecular tracer.  For example, the
ammonia mapping of B5 in Perseus by \citet{Pineda2010} revealed a
sharp decrease in line width in the central 0.1~pc.  This is thought
to be the result of the decay of supersonic turbulence present on
larger scales; the observed line widths are subsonic for a
thermal/kinetic temperature of 10~K.  \citet{Foster2009} present
results from an ammonia survey of many other cores in Perseus and find
them mostly quiescent ($\sigma_{\rm nonthermal} \approx
0.12$~km~s$^{-1}$, but see their histograms for details of the full
survey).  The modeling of two lines gives kinetic temperatures about
12~K; derived excitation temperatures are lower \citep{Foster2009}.
The dust temperature from the submillimetre spectral energy
distribution observable by \emph{Herschel} is typically low (down to
10~K) in the high column density regions in low mass star forming
regions (see the temperature maps in, e.g., \citealp{Bontemps2010} and
\citealp{Arzoumanian2011}).  Coupling of the kinetic temperature in
the gas to the dust temperature comes into play for $n_{\rm H2} >
10^4~{\rm cm}^{-3}$ \citep{Goldsmith2001}.  According to
Fig.~\ref{fig_centraldensity}, this becomes relevant for strongly
self-gravitating cylinders ($f_{\rm cyl} > 0.7$).  Taken together,
while perhaps a platonic ideal, a small dense isothermal core with $K$
corresponding to 10~K is reasonable as a reference value.  However,
$K$ might be higher than this in some environments or stages of
evolution.  And of course the structure might not be exactly
isothermal.

\section{\label{app1}Mean column density of spheres}

In paper~II the approximation for the mean column densities of
cylinders and spheres as a function of the overpressure were
presented.  Here, we want to comment on the values for spheres and
give an exact expression for the asymptotic behaviour at large
overpressures.

The mass of a spherical isothermal self-gravitating cloud which is in
pressure equilibrium with the surrounding medium ($p(z_{\rm
  cl})=p_{\rm ext}$) is given by (paper~I)
\begin{equation}
	\label{eq_mass}
	M_{\rm sph} = \frac{K^2}{\sqrt{4\pi G^3 p_{\rm ext}}} e^{-w(z)/2} \int_0^{z}{\rm d}z'\,z'^2\,e^{-w(z')}.
\end{equation}
Here, $w=\phi/K=\ln{p_c/p_{\rm ext}}$ is the unit free potential and
$z=r A$ the unit free radius of the Lane-Emden equation
\begin{equation}
	\frac{1}{z^2}\frac{{\rm d}}{{\rm d}z}\left(z^2\frac{{\rm d}\omega}{{\rm d}z}\right)=e^{-\omega},
\end{equation} 
where $A^2 = 1/r_0^2=4\pi G\rho_{\rm c} / K$.
Using the expression for the mass (Eq.~\ref{eq_mass}) and
\begin{equation}
	r_{\rm cl} =z_{\rm cl}/A =  z_{\rm cl} \frac{K}{\sqrt{4\pi G p_{\rm ext}}}e^{-\omega(z_{\rm cl})/2},
\end{equation}
for the cloud radius we obtain for the mean mass surface density of an
isothermal self-gravitating sphere
\begin{equation}
	\left<\Sigma\right>(z_{\rm cl})  =    \frac{M_{\rm sph}(z_{\rm cl})}{\pi r_{\rm sph}^2(z_{\rm cl})}
					 =  \frac{\sqrt{4\pi G p_{\rm ext}}}{\pi G} \zeta(z_{\rm cl}), 
\end{equation}
where
\begin{equation}
	\label{eq_zeta}
		\zeta(z_{\rm cl}) = \frac{e^{-\omega(z_{\rm cl})/2}\int_0^{z_{\rm cl}}{\rm d}z
					\,z^2\,e^{-\omega(z)}}{z_{\rm cl}^2\,e^{-\omega(z_{\rm cl})}}.
\end{equation}

We want to analyze the function $\zeta(z)$ for the values $z_i$ with
$i=1,2,...$ at which a spherical cloud of a fixed mass produces pressure
maxima at the cloud outskirts for changing cloud size $z_{\rm cl}$
\citep{Fischera2008}.  The global pressure maximum at $z_1=z_{\rm
  max}$ with $z_{\rm max}\approx 6.451$ characterizes the critical
stable cloud which has an overpressure $e^{\omega(z_{\rm max})}\approx
14.04$. The last two terms in Eq.~\ref{eq_mass} for a critical stable
cloud produce a value of approximately 4.191.  We obtain the condition
for the pressure maxima from Eq.~\ref{eq_mass} which we transform to
an expression for the pressure at the cloud outskirts $p(z_{\rm cl})$.
At the pressure maxima we find that the derivative of the potential
needs to fulfill the condition
\begin{equation}
	\label{eq_condition}
	\left.\frac{{\rm d}\omega}{{\rm d}z}\right|_{z_i} = 2 \frac{z_i^2\,e^{-\omega(z_i)}}{\int_0^{z_i}{\rm d}z\,z^2\,e^{-\omega(z)}}.
\end{equation}
On the other hand, from the Lane-Emden equation it follows that
\begin{equation}
	z^2\frac{{\rm d}\omega}{{\rm d}z} = \int_0^{z}{\rm d}z'\,z'^2\,e^{-\omega(z')}.
\end{equation}
Using this expression to replace the derivative in
Eq.~\ref{eq_condition} we obtain for the condition for the pressure
maxima
\begin{equation}
	\label{eq_condition2}
	\int_0^{z_i} {\rm d}z\,z^2\,e^{-\omega(z)} = \sqrt{2} z^2_i\,e^{-\omega(z_i)/2}.
\end{equation}
For clouds producing maxima at the cloud outskirts the two last terms
in Eq.~\ref{eq_mass} of the cloud mass are therefore equal to
$\sqrt{2}\,z_{\rm i}^2\,e^{-\omega(z_{\rm i})}$. For those clouds we
get the alternative mass equation
\begin{equation}
	M_{\rm sph}(z_i) = \frac{K^2}{\sqrt{4\pi G^3 p_{\rm ext}}} \sqrt{2}\,z_i^{2}\,e^{-\omega(z_i)}
\end{equation}
For the critical stable cloud we have the condition $\sqrt{2}z^2_{\rm
  max}\,e^{-\omega(z_{\rm max})}\approx 4.191$ with $z_{\rm
  max}\approx 6.451$.

If we insert the condition \ref{eq_condition2} in Eq.~\ref{eq_zeta} we
see that the pressure maxima are at $z$-values where $\zeta(z_i) =
\sqrt{2}$ and we obtain as the asymptote for large overpressure:
\begin{equation}\label{eq_meansigma}
	\left<\Sigma\right>   =  \frac{\sqrt{4\pi G p_{\rm ext}}}{\pi G} \sqrt{2}.
\end{equation}
The asymptotic value of the mean column density given in
Eq.~\ref{eq_meancolumn} is obtained using the relation
\begin{equation}
	\left<N_{\rm H}\right> = \left<\Sigma\right> (\nrho)^{-1} .
\end{equation}

\section{\label{sect_fwhmspheres}The fwhm of isothermal
  self-gravitating clouds}


To derive the \FWHM{} we replaced the density through
$\rho(r)=\rho_{\rm c}e^{-\omega}$ and the radius by $r=z/A$. In case
of cylinders the potential is given by $\omega(z)=2\ln\{1+z^2/8\}$
(Eq.~\ref{eq_densityprofile}).  The mass surface density at impact
parameter $\bar z \le z_{\rm cl}$ is then given by
\begin{equation}
	\Sigma(\bar z) = 2 \sqrt{\frac{p_{\rm ext}}{4\pi G}} e^{\omega(z_{\rm cl})/2}\int_{\bar z}^{z_{\rm cl}}{\rm d}z\,
		\frac{z}{\sqrt{z^2-\bar z^2}} e^{-\omega(z)}.
\end{equation}
The impact parameter $z_{\FWHMEQ}$ for the \FWHM{} for an isothermal pressurized cloud is determined by
\begin{equation}
 	2\int_{z_{\FWHMEQ}}^{z_{\rm cl}}{\rm d}z\frac{z}{\sqrt{z^2-z_{\FWHMEQ}^2}}e^{-\omega(z)} 
		= \int_{0}^{z_{\rm cl}}{\rm d}z\,e^{-\omega(z)}.
\end{equation}

For $z_{\rm cl}\ll 1$ the density profile is essentially flat and we
obtain the simple asymptotic behaviour $z_{\FWHMEQ} = \sqrt{3/4}z_{\rm
  cl}$ which gives
\begin{equation}
\label{fwhmandsigma0}
	\FWHMEQ(z_{\rm cl}\ll 1) = \frac{\sqrt{3}}{2} \frac{K}{p_{\rm ext}}\Sigma(0).
\end{equation}
At sizes $z_{\rm cl}\gg 1$ the right hand side approaches
asymptotically a constant value.  For spheres we get
\begin{equation}
	I_{\infty} = \int_0^{\infty}{\rm d}z\,e^{-\omega(z)} \approx 3.028.
\end{equation}
For cylinders we obtain
\begin{equation}
	I_{\infty} = \int_{0}^{\infty} {\rm d}z\,\frac{1}{(1+z^2/8)^2} = \frac{\pi}{\sqrt{2}}.
\end{equation}

This constant determines the impact parameter for high overpressure or
large $z_{\rm cl}$. For spheres we find
\begin{equation}
	z_{\FWHMEQ}  \approx 2.997
\end{equation}
and for cylinders
\begin{equation}
	z_{\FWHMEQ} = \sqrt{8(2^{2/3}-1)}.
\end{equation}

The relation $r=z/A$ provides
\begin{equation}
	\FWHMEQ(z_{\rm cl}\gg 1) = 2 z_{\FWHMEQ}\frac{K}{\sqrt{4\pi G p_{\rm ext}}} e^{-\omega(z_{\rm cl})/2}.
\end{equation}
As noted in Sect.~\ref{sect_fwhm}, in the limit of high overpressure
($z_{\rm cl}\gg 1$) the \FWHM{} for constant $T$ and $p_{\rm ext}$
decreases inversely proportional to the square root of the overpressue
$p_{\rm c}/p_{\rm ext}$. Replacing the central pressure by the mass
surface density through the centre gives finally for $z_{\rm cl}\gg 1$
the asymptotic behaviour
\begin{equation}
\label{fwhmandsigma1}
	\FWHMEQ(z_{\rm cl}\gg 1) = \frac{z_{\FWHMEQ}}{\pi} I_{\infty} \frac{K}{G}\frac{1}{\Sigma(0)},
\end{equation}
so that $\FWHMEQ\propto K\Sigma^{-1}(0)$. For high overpressure
the relation \FWHMEQ-$\Sigma(0)$ does not depend on the external pressure.


\section{\label{sect_appC}On combining power laws}

Here, we discuss the main properties of the function used to fit the
relation of the \FWHM{} and the central column density $N_{\rm
  H}(0)$. Call these $y$ and $x$, respectively.  There are asymptotic
limits relating these for small and large mass ratios (e.g., the power
laws in Eq.~\ref{fwhmnh0} and Eq.~\ref{fwhmnh1} for cylinders).  We
assume that two power laws
\begin{eqnarray}
	y_1 &=& \xi_1 x^{a-1}, \quad x\ll \bar x,\\
	y_2 &=& \xi_2 x^{-b-1},\quad x\gg \bar x,
\end{eqnarray}
with $a>1$ and $b>-1$, can be combined in an interpolating formula
\begin{equation}
	y(t) = \xi_1\bar x^{a-1} \frac{t^{a-1}}{(1+(t/t_0)^\gamma)^{\frac{a+b}{\gamma}}},
\end{equation}
where we have introduced $t=x/\bar x$ and where
$t_0=(\xi_2/\xi_1)^{1/(a+b)}/\bar x$.  As can be verified, the equation
provides the correct limits for $t\ll t_0$ and $t\gg t_0$.  The
parameter $\gamma$ describes the smoothness of the transition and is
adjusted to optimize the fit.  For $\gamma\rightarrow\infty$ the
function becomes a simple broken power law.  The maximum lies at
\begin{equation}
	t_{\rm max}=t_0\sqrt[\gamma]{\frac{a-1}{b+1}},
\end{equation}
with
\begin{equation}
	y_{\rm max} = \sqrt[a+b]{\xi_1^{b+1}\xi_2^{a-1}}\sqrt[\gamma]{\frac{(a-1)^{a-1}(b+1)^{b+1}}{(a+b)^{a+b}}}.
\end{equation}
If we identify $\bar x$ as the value where the function $y$ reaches
its maximum we have
\begin{eqnarray}
	t_0 & = & \sqrt[\gamma]{\frac{b+1}{a-1}},\\
	\bar x &=& \sqrt[a+b]{\xi_2/\xi_1} \sqrt[\gamma]{\frac{a-1}{b+1}}.
\end{eqnarray}
We find that the combined function can be written as
\begin{equation}
	y(t) = y_{\rm max} C \frac{(t/t_0)^{a-1}}{(1+(t/t_0)^\gamma)^{(a+b)/\gamma}},
\end{equation}
with
\begin{equation}
	C^{-1} = \sqrt[\gamma]{\frac{(a-1)^{a-1}(b+1)^{b+1}}{(a+b)^{a+b}}}.
\end{equation}
In the special case $a=2$, $b=0$ as for the relation between \FWHM{}
and $\Sigma(0)$ for cylinders or in good approximation for a sphere we
have
\begin{eqnarray}
	t_0 &=& 1, \\
	y_{\rm max} &=& \sqrt{\xi_1 \xi_2}/C,\\
	C &=& \sqrt[\gamma]{4},\\
	\bar x &=& \sqrt{\xi_2/\xi_1}.
\end{eqnarray}

\section{Analytic approximations of the length and time scales of perturbations}\label{Sect_polapprox}

The derived dependence of the perturbation lengths $\lambda_{\rm cr}$
and $\lambda_{\rm m}$ and the initial growth time $\tau_{\rm m}$ on the
mass ratio and the magnetic field strength are based on the work of
\citet{Nagasawa1987} who presented calculations for $f_{\rm cyl}=0.01$,
0.5, and 1.0.  We assumed that the mass ratio dependence of the time
scale given in units $1/\sqrt{4\pi G \rho_{\rm c}}$ and the ratio of the
disturbance length to the \FWHM{} of the cylinder for non-magnetized
cylinders can be described through a fourth-order polynomial function
\begin{equation}
	y(x) = \sum\limits_{i=0}^{4}a_i x^i,
\end{equation}
in $x = f_{\rm cyl}$ with vanishing derivatives for low and high mass
ratios ($y'(0) = y'(1)=0$). The curves reproduce the theoretical values
at $f_{\rm cyl}=0$, 0.5, and 1.0.  The disturbance lengths can be
transferred to absolute lengths using a fifth-order polynomial
approximation for the \FWHM{} in units of the characteristic length
$\sqrt{8}r_0$ given by
\begin{equation}
	\FWHMEQ_{\rm cyl}/(\sqrt{8}r_0) = \sum\limits_{i=1}^{5} a_i f_{\rm cyl}^{i/2}.
\end{equation}
The approximation is $1\%$ accurate and provides the correct asymptotes
at low and high $f_{\rm cyl}$. The coefficients $a_i$ for the polynomial
representations are given in Table~\ref{table_polynom}.

To estimate the dependence of the magnetic field on the parameters
$\lambda_{\rm cr}$, $\lambda_{\rm m}$, and $\tau_{\rm m}$ we used at low
mass ratios the dispersion relation for an incompressible
self-gravitating cylinder which has been shown to be a valid
representation for $f_{\rm cyl}\rightarrow 0$. The relation is given by
\citep{Nagasawa1987}
\begin{equation}
	\frac{\omega^2}{4\pi G\rho_{\rm c}} = \frac{\xi I_{1}(\xi)}{I_0(\xi)}\left[K_0(\xi)I_0(\xi)-\frac{1}{2}\right]-\frac{p^2}{8f_{\rm cyl}}
	\frac{\xi}{I_{0}(\xi)K_1(\xi)},
\end{equation}
where $I_\nu(\xi)$ and $K_\nu(\xi)$ are modified Bessel functions, $p^2 = B^2/(4\pi \rho_c K)$,
$\xi = (2\pi/\lambda) R$, and $\omega = 1/\tau$.  The behaviour at
larger $f_{\rm cyl}$ we described through a fourth-order polynomial
function where we have chosen $x=f_{\rm cyl}^{1/3}$ for $\tau_{\rm m}$
and $x = f_{\rm cyl}^{1/2}$ for the disturbance lengths $\lambda_{\rm
  cr}$ and $\lambda_{\rm m}$. To the analytical function at low mass
ratios we produced a smooth transition and assumed a vanishing
derivative at $f_{\rm cyl}=1$.  The curves were chosen to reproduce the
values of \citet{Nagasawa1987} at $f_{\rm cyl}=0.5$ and $f_{\rm cyl}=1$.
For strong magnetic fields the left boundary for the polynomial
approximation is set to a value were we obtain at high mass ratios
essentially a flat curve. The corresponding curves only show
qualitatively at which $f_{\rm cyl}$ the parameters are affected by the
magnetic field.

\begin{table}[htbp]
	\caption{\label{table_polynom} Constants of polynomial approximations}
	\begin{tabular}{lccccccc}
				& $a_0$ & $a_1$ & $a_2$ & $a_3$	& $a_4$  & $a_5$\\
		$\tau_{\rm m}\sqrt{4\pi G\rho_{\rm c}}$ & 4.08 & 0.00 & -2.990 & 1.460 & 0.400 & 0.00\\
		$\lambda_{\rm cr}/\FWHMEQ_{\rm cyl}$ & 3.39 & 0.00 & -2.414 & 1.588 & 0.016 & 0.00\\
		$\lambda_{\rm m}/\FWHMEQ_{\rm cyl}$ & 6.25 & 0.00 & -6.890 & 9.180 & -3.440 & 0.00\\
		$\frac{\FWHMEQ_{\rm cyl}}{\sqrt{8}r_0}$ & 0.00 & 1.732 & 0.000 & -0.041 & 0.818 & -0.976
	\end{tabular}
\end{table}


\end{document}